\documentclass{article}
\usepackage[11pt]{extsizes}
\usepackage{ulem}
\usepackage[english]{babel}
\usepackage[T1]{fontenc}
\usepackage{listings}
\usepackage{amsmath}
\usepackage{mathrsfs}
\usepackage{hyperref}
\usepackage{ifthen}
\usepackage{blindtext}
\usepackage{mathtools}
\usepackage{amsmath}
\usepackage{amssymb}
\usepackage{url}
\usepackage{textcomp}
\usepackage{dsfont}
\usepackage{bbm}
\usepackage{verbatim}
\usepackage{glossaries}
\usepackage{appendix}
\usepackage{caption}
\usepackage{subcaption}
\usepackage[square,sort,comma,numbers]{natbib}
\usepackage{xcolor}

 \usepackage{array}
\usepackage{booktabs}
\usepackage{tabularx}
\usepackage{color}
\definecolor{mygreen}{RGB}{28,172,0} 
\definecolor{mylilas}{RGB}{170,55,241}
\newcommand{\R}{\mathbb{R}}
\newcommand{\LL}{\mathbf{\Lambda}}

\newcommand{\N}{\mathbb{N}}
\newcommand{\E}{\mathbb{E}}

\newcommand{\Prob}{\mathbb{P}}

\newcommand{\Tau}{\mathcal{T}}
\usepackage{mathtools}

\DeclarePairedDelimiterX{\expectarg}[1]{(}{)}{%
  \ifnum\currentgrouptype=16 \else\begingroup\fi
  \activatebar#1
  \ifnum\currentgrouptype=16 \else\endgroup\fi
}

\newcommand{\innermid}{\nonscript\;\delimsize\vert\nonscript\;}
\newcommand{\activatebar}{%
  \begingroup\lccode`\~=`\|
  \lowercase{\endgroup\let~}\innermid 
  \mathcode`|=\string"8000
}

\newtheorem{theo}{Theorem}

\newtheorem{prop}{Proposition}
\newtheorem{lem}{Lemma}
\newtheorem{remark}{Remark}

\usepackage{graphicx}
\usepackage[utf8]{inputenc}
\usepackage{mathenv}
\usepackage{gensymb}
\usepackage{subcaption}
\usepackage{tikz}
\usetikzlibrary{shapes}
\usetikzlibrary{trees}
\usetikzlibrary{calc}
\newcommand\tab[1][0.5cm]{\hspace*{#1}}
\usepackage[top=2cm, bottom=2cm, left=2cm, right=2cm]{geometry}
\usepackage{relsize}
\usepackage{multirow}
\usepackage{lipsum}
\usepackage{bigints}
\usepackage{amsmath}               

\usepackage[section]{placeins}
\usepackage{caption}
\usepackage{float}
\usepackage[justification=centering]{caption}
\usepackage{bm}

\usepackage{mathtools, stmaryrd}
\usepackage{xparse} \DeclarePairedDelimiterX{\Iintv}[1]{\llbracket}{\rrbracket}{\iintvargs{#1}}
\NewDocumentCommand{\iintvargs}{>{\SplitArgument{1}{,}}m}
{\iintvargsaux#1} %
\NewDocumentCommand{\iintvargsaux}{mm} {#1\mkern1.5mu..\mkern1.5mu#2}

\hypersetup{
   colorlinks=true, 
   breaklinks=true, 
   urlcolor= blue, 
   linkcolor= blue,
   pdfcreator  = {\LaTeX},
   pdfproducer = {Kile}
}

\DeclareMathOperator*{\argmin}{arg\,min} 

\usepackage{natbib}
\usepackage{anysize}

\marginsize{2cm}{2cm}{2cm}{2cm}

\usepackage{anyfontsize}

\usepackage{appendix}

\usepackage{listings} 
\definecolor{dkgreen}{rgb}{0,0.6,0} 
\definecolor{gray}{rgb}{0.5,0.5,0.5} 
\usepackage{fancyhdr} 
\usepackage{amssymb}
\usepackage{enumitem}

\usepackage{authblk}
\title{\huge A Nested Factor Model for Equity Markets: Reconciling Multifractal Stock Returns and Rough Index Volatilities}

\author[1]{Othmane Zarhali}
\author[2,3]{Cecilia Aubrun}
\author[1]{Emmanuel Bacry} 
\author[2,4,5]{Jean-Philippe Bouchaud}
\author[6]{Jean-François Muzy}
\affil[1]{Ceremade, CNRS-UMR 7534, Université Paris-Dauphine PSL, Place du Maréchal de Lattre de Tassigny, 75016 Paris, France}
\affil[2]{Chair of Econophysics and Complex Systems, École polytechnique, 91128 Palaiseau Cedex, France}
\affil[3]{LadHyX UMR CNRS 7646, École polytechnique, 91128 Palaiseau Cedex, France}
\affil[4]{Capital Fund Management, 23 Rue de l’Université, 75007 Paris, France}
\affil[5]{Académie des Sciences, 23 Quai de Conti, 75006 Paris, France}
\affil[6]{SPE CNRS-UMR 6134, Université de Corse BP 52, 20250 Corte, France}

\date{}
\hypersetup{
    colorlinks=False,
    linkcolor=black,
    citecolor=black,
    urlcolor=blue
}

\newcounter{subsubsubsection}[subsubsection]

\makeatletter
\renewcommand\paragraph{\@startsection{paragraph}{4}{\z@}%
  {3.25ex \@plus1ex \@minus.2ex}%
  {1em}%
  {\normalfont\normalsize\bfseries}}
\renewcommand\subparagraph{\@startsection{subparagraph}{5}{\parindent}%
  {3.25ex \@plus1ex \@minus .2ex}%
  {-1em}%
  {\normalfont\normalsize\bfseries}}
\makeatother

\begin{document}

\maketitle
\begin{abstract} 
The Nested factor model was introduced by Chicheportiche\textit{ et al.} in \cite{chicheportiche2015nested} to represent non-linear correlations between stocks.  Stock returns are explained by a standard factor model, but the (log)-volatilities of factors and residuals are themselves decomposed into factor modes, with a common dominant volatility mode affecting both market and sector factors but also residuals. Here, we consider the case of a single factor where the only dominant log-volatility mode is rough, with a Hurst exponent $H \simeq 0.11$ and the log-volatility residuals are ``super-rough'' or ``multifractal'', with $H \simeq 0$. We demonstrate that such a construction naturally accounts for the somewhat surprising stylized fact reported by Wu\textit{ et
al.} in \cite{wu2022rough}, where it has been observed that the Hurst exponents of stock {\it indexes} are large compared to those of individual stocks. We propose a statistical procedure to estimate the Hurst factor exponent from the stock returns dynamics together with theoretical guarantees of its consistency. We demonstrate the effectiveness of our approach through numerical experiments and apply it to daily stock data from the S\&P500 index. The estimated roughness exponents for both the factor and idiosyncratic components validate the assumptions underlying our model.
\end{abstract}

\newpage
\section{Introduction}

Volatility modelling has greatly evolved since the Bachelier model in 1900 \cite{bachelier1900theorie}, where prices follow a Brownian motion with constant volatility. It has long been recognized that such a model fails to capture the salient features that justify the very existence of option markets. Indeed, volatility being fluctuating and uncertain, volatility {\it bets} become relevant. Furthermore, price jumps and fat tails prevent perfect hedging strategies, so that option markets are non-redundant and allow real risk to be exchanged between participants. 

In addition, volatility fluctuations have special statistical properties, as volatility clustering. As noted long ago by Mandelbrot \cite{mandelbrot1997variation}, {\it large price changes of either sign tend to be followed by large price changes, and small price changes tend to be followed by small price changes}. Interestingly, one observes periods of high (/low) volatilities at all frequencies, from minutes to years. Such a remarkable ``long-memory'' phenomenon calls for models with specific mathematical properties.

One family of such models that has become extremely popular in the last 25 years assumes that the log-volatility process  is a fractional (Gaussian) Brownian motion $B^H$, characterized by a Hurst exponent $H \in  (0,1)$ that describes the ``roughness'' of its trajectory through
\begin{equation}\label{eq:fBM_def}
    \mathbb{E}\left((B_t^H-B_s^H)^2\right) = (t-s)^{2H}.
\end{equation} 
The case $H=\frac{1}{2}$ corresponds to the classic Brownian motion. Models with $H<\frac{1}{2}$ corresponds to ``rough'' log-volatility dynamics, whereas $H>\frac{1}{2}$ corresponds to log-volatilities smoother than Brownian motion. For instance, pioneering work by Comte and Renaud in 1998 \cite{comte1998long} proposed a smooth volatility process with $H> \frac{1}{2}$ to account for the long-range memory of volatility. At the other extreme, two of us (E.B. and J.-F. M.) introduced in 2000 the Multifractal Random Walk (MRW) model \cite{bacry2001multifractal,muzy2000modelling,muzy2002multifractal}, which formally corresponds to $H\xrightarrow[]{}0^+$. Later, this multifractal regime was also observed in real data \cite{muzy2000modelling, bacry2001modelling, gatheral2018volatility, bolko2020roughness,fukasawa2019volatility}, as clarified recently in \cite{wu2022rough} (more on this later, and see also \cite{neuman2018fractional}). 

Rough volatility modelling has gathered considerable momentum since 2016, after the seminal paper of Gatheral \textit{et al.} \cite{gatheral2018volatility}, followed by a flurry of papers in which new mathematical tools are developed to price options within such a framework \cite{gatheral2018volatility,bayer2016pricing,fukasawa2021volatility,livieri2018rough,bayer2023rough}. At variance with the ``multifractal'' $H\xrightarrow[]{}0^+$ MRW model, Gatheral et al. found that the roughness index $H$ of the log-volatility of major stock indexes (such as the S\&P500 and the DAX) is around $0.14$ (see also \cite{bennedsen2022decoupling}).

On the other hand, factor models in asset pricing have been central to the empirical investigation of risk premia and portfolio returns. The canonical framework, the Capital Asset Pricing Model (CAPM) of \cite{sharpe1964capital} and \cite{lintner1965security}, posits a single market factor as the driver of expected returns. However, empirical evidence soon revealed systematic deviations from CAPM predictions, motivating the development of multifactor models such as the Fama-French three-factor model (see \cite{fama1993common}). Beyond these linear specifications, modern approaches exploit large panels of test assets and apply statistical methods such as principal component analysis and sparse estimation to extract latent risk factors \cite{connor1988risk, bai2002determining}. More recent advances integrate economic theory with machine learning techniques to improve factor selection and to address the proliferation of proposed anomalies (see for instance \cite{harvey2016cross}).
\\

\begin{figure}
    \centering
    \includegraphics[width=0.7\linewidth]{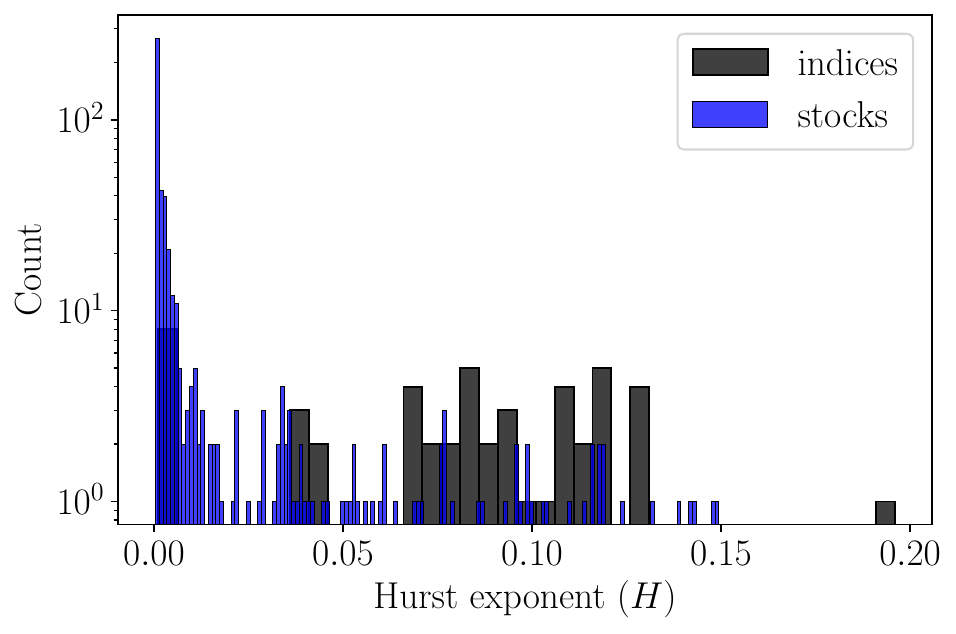}
    \caption{Hurst exponents estimation of 503 stocks and 49 indexes using the calibration method of Wu \textit{et al.} \cite{wu2022rough} on daily Garman-Klass estimation of the log-volatility from 2013 to 2023.}
    \label{fig:wu_empirical_results}
\end{figure}

Seeking to build a unifying framework that conciliates the observed (small) Hurst exponent of order $H\sim0.1$ with the asymptotic behavior where $H\xrightarrow[]{}0^+$, P. Wu, E. Bacry and J.-F Muzy introduced in 2022 the logarithmic Stationary fractional Brownian Motion (Log S-fBM) \cite{wu2022rough} (see also \cite{forde2022riemann,fyodorov2016fractional,hager2022multiplicative} for alternative approaches).  Calibrating their model on real world data led to intriguing results. While the roughness index of stock indexes is compatible with the values reported by Gatheral et al. ($H \approx 0.1$), individual stocks exhibit significantly greater roughness ($H\approx0$) as illustrated in Figure \ref{fig:wu_empirical_results} (see also Figure 10 of \cite{wu2022rough}).

Such a result is puzzling and at first sight paradoxical, since a stock index is nothing but the average of individual stocks -- how can the volatility of the latter be {\it smoother} than that of all its constituents? This observation is the motivation for the present paper, where we revisit the ``Nested Factor Model'' (NFM) of R. Chicheportiche and one of us (J.-P.B) \cite{chicheportiche2015nested} and show that such a framework indeed allows us to reconcile all empirical findings.   


In a nutshell, our basic intuition is that the roughness exhibited of a composite signal, composed of several signals with varying degrees of roughness, tends to be dominated by the roughest component. Consequently, the empirical observations of Figure~\ref{fig:wu_empirical_results}, suggests that the ``multifractal'' volatility contributions to individual stocks should nearly disappear when averaging over single stocks and only leave the common, smoother volatility contribution to emerge at the index level. The rougher component, characterized by  $H\approx 0$, is thus expected to be associated with the idiosyncratic dynamics of individual stocks. 

Factor models offer a natural framework for investigating such a hypothesis. Indeed, thinking of stock returns as the sum of a factor component (common to all stocks) and an idiosyncratic component (specific to each stock), it is clear that if the idiosyncratic component is rougher than the factor component, it will dominate at the stock level and disappear at the index level. However, things turn out to be a little more complex since it has been shown in several previous studies that the volatility of the idiosyncratic component is strongly correlated with the volatility of the index, see \cite{cizeau2001correlation,allez2011individual, chicheportiche2012joint}. This is precisely the motivation for the Nested factor model (NFM) proposed in \cite{chicheportiche2015nested}, which accurately captures the anomalous copula structure between pairs of stocks, that reflects the existence of a common volatility mode in their idiosyncratic component.  

The outline of the article is as follows. Section~\ref{sec:models_of_interest} presents the models we are interested in, i.e. the Log S-fBM and the Nested Factor model. Section~\ref{sec:model} introduces our framework which couples the NFM with SfBM as log-volatility modes, and that we thereafter call the Nested Log S-fBM factor model (N-SfFM). The rest of the paper aims at verifying that the N-SfFM replicates the empirical findings of Wu \textit{et al.} exposed in Fig.~\ref{fig:wu_empirical_results}. We intend to do so analytically, numerically and on real data. Specifically, Section~\ref{sec:conditions_on_parameters} determines the limits of the different regimes by specifying conditions on the model parameters. Finally, Section~\ref{Section:numerics} presents our numerical  experiments, while Section~\ref{Section:empirical} summarizes our empirical results.


\section{Theoretical Backdrop}
\label{sec:models_of_interest}

\subsection{The log Stationary fractional Brownian Motion}
\label{sec:logSfBM}


In this Section, we outline the main characteristics of the Log S-fBM model introduced in \cite{wu2022rough} as a consistent interpolation between rough models with $0 < H < \frac12$ and the ``multifractal'' MRW, that corresponds to $H\xrightarrow[]{}0^+$. For more comprehensive details, we refer the reader to \cite{wu2022rough}.

For any Hurst exponent $0<H<\frac{1}{2}$, the Stationary fractional Brownian Motion (S-fBM) $(\omega_{H,T}(t))_t$ is a stationary Gaussian process\footnote{Notice that all the framework introduced in this work can be naturally extended to a wide class of infinitely divisible laws along the same line as discussed in \cite{muzy2002multifractal,bacry2003log,wu2022rough}.} defined by its covariance function:
\begin{equation}
    C_{\omega_{H,T}}(\tau) = \text{cov}(\omega_{H,T}(t),\omega_{H,T}(t+\tau))\\
    = \begin{cases}
        \frac{\nu^2}{2}\left(1-(\frac{\tau}{T})^{2H} \right),\quad \text{ when } |\tau|<T\\
        0 \quad \text{otherwise}
    \end{cases}
    \label{eq:covSfBM}
\end{equation}
where $\nu$ is a dimensionless parameter that controls the amplitude of the process, and $T$ is the time range of correlations.  The mean value of $\omega_{H,T}$ is chosen  such a way that:
\begin{equation}
\label{eq:mean_omega}
 \mathbb{E}\left(e^{\omega_{H,T}(t)} \right) = 1 \; .
\end{equation}
The Log S-fBM random measure $(M_{H,T}(t))_t$ is then defined as:
\begin{equation}\label{eq:NFM_MRM_M_def}
    {\rm d}M_{H,T}(t) = \exp(\omega_{H,T}(t)){\rm d}t.
\end{equation}
Wu \textit{et al.} primarily demonstrate that as $H$ goes to 0 and $\nu$ tends toward infinity while keeping fixed the so-called dimensionless intermittency coefficient:
\begin{equation} 
\label{eq:rel_nu_lambda2}
\lambda^2=H(1-2H)\nu^2,
\end{equation}
the Log S-fBM $(M_{H,T}(t))_t$ converges toward a Multifractal Random Measure (MRM) (that will be consequently noted $M_{0,T}$) previously introduced in \cite{bacry2001multifractal,bacry2001modelling,muzy2000modelling,muzy2002multifractal}. Even though a vanishing Hurst exponent leads to an ill defined S-fBM process because of the exploding variance (see Eq.~\eqref{eq:covSfBM} and Eq.~\eqref{eq:rel_nu_lambda2}, the limit (in the weak sense) when the Hurst vanishes is in the level of the Log S-fBM random measure (Eq.~\eqref{eq:NFM_MRM_M_def}) that converges toward $M_{0,T}$.

The MRM is classically defined as the limit measure:
\begin{equation}\label{eq:NFM_MRM_M0_def}
    {\rm d}M_{0,T}(t) = \lim_{l \rightarrow 0^+} \exp(\tilde \omega_{l,T}(t)){\rm d}t
\end{equation}
where $\tilde \omega_{l,T}$ is a stationary Gaussian process, whose mean is chosen such that
$ \mathbb{E}(e^{\tilde \omega_{l,T}(t)}) = 1$, and whose covariance is given by:
\begin{eqnarray}
\label{eq:logcorrelprocesswithcutoff}
    \begin{aligned}
    C_{\tilde \omega_{l,T}}(\tau) =\text{cov}(\tilde \omega_{l,T}(t),\tilde \omega_{l,T}(t+\tau)) =  \begin{cases}
        -\lambda^2 (\ln({l}/{T}) - 1 + \tau/l),\quad \text{ when } |\tau|<l\\
        -\lambda^2 \ln({\tau}/{T}),\quad \text{ when } l< |\tau|<T\\
        0 \quad \text{otherwise}
    \end{cases}.
\end{aligned}
\end{eqnarray}



Hence, this framework harmonizes the cases $H>0$ and $H \xrightarrow{}0$. 

Wu \textit{et al.} complete their model by proposing financial price dynamics to be described with the Log S-fBM, such that return of an asset ${\rm d}x_t$ writes:\footnote{Let us note that this equation can be problematic when $H\rightarrow 0$. In that case, the process is defined through a subordination of the Brownian motion $B$.}
\begin{equation*}
    {\rm d}x_t = \sqrt{\frac{{\rm d}M_{H,T}(t)}{{\rm d}t}}\,{\rm d}B_t 
\end{equation*}

where $(M_{H,T}(t))_{t\geq 0}$ is a Log S-fBM measure and $(B_t)_{t}$ is a standard Brownian motion.

Within this framework, Wu \textit{et al.} introduced two Generalized Methods of Moments to calibrate the Log S-fBM on empirical data. These methods rely on estimating the autocovariance of ${\rm d}M_{H,T}$ or $\ln\left({\rm d}M_{H,T}\right)$, for a fixed set of lags (see \cite{wu2022rough}, Proposition 6). By applying their calibration method to daily volatility  Garman-Klass estimates, results displayed in Figure~\ref{fig:wu_empirical_results} were obtained (see also Figure 10 in \cite{wu2022rough}). Notably, the discrepancy in the order of magnitude between the Hurst exponents of indexes (in black) and individual stocks (in blue) is striking. Indeed, as noted in the introduction above, stock log-volatilities are rougher ($H\approx0$) than that of indexes $(H\approx0.1)$.
Empirical observations suggest that the correlation scale \( T \) is quite large (see as a support the experiments of Figures 8 and 9 in \cite{MuzyBaileBacry2013_PRE}). As highlighted in \cite{wu2022rough}, when the total sample size \( L \) is smaller than \( T \), the statistical properties of the Log S-fBM are indistinguishable from those observed when \( T = L \). Therefore, in what follows, we assume that \( T \) corresponds to the full sample size, typically spanning a few years.

\subsection{The Nested Factor Model}

Introduced in \cite{chicheportiche2015nested}, the Nested Factor model (NFM) seeks to replicate the dynamics of stocks returns and their interplay with market dynamics. The development of this model stemmed from the observation that \textit{joint distribution of stock returns is not elliptical} \cite{chicheportiche2012joint}.

In this context, we define the Nested Factor Model with a single factor ($M=1$) and a single volatility mode. The extension to $M$ factors is outlined in Appendix~\ref{app:MfactorNFM}. This model is applied to the returns,  ${\rm d}x^{i}_t$ of stock $i$ at time $t$ as follows:

    \begin{equation}\label{eq:NFM_definition}
   {\rm d}x^{i}_t = \beta_i \, {\rm d}f_t + {\rm d}\epsilon^{i}_t \quad \text{with} \quad 
   \begin{cases}
        {\rm d}f_t = e^{\frac{\Omega_t}{2}} \, {\rm d}W^0_t, \\
        {\rm d}\epsilon^{i}_t = \sigma_i \, e^{\frac{\tilde{\omega}^{i}_t}{2}} \, {\rm d}B^{i}_t,
    \end{cases}
    \quad \text{and} \quad \tilde{\omega}^{i} := \gamma_i \Omega + \omega^{i}.
\end{equation}
where 
\begin{itemize}
\setlength\itemsep{0.01em}
    \item The time series $f_t$ represents the market itself, and $({\rm d}f_{t})_t$ can be accurately proxied by the returns of the S\&P500 index. 
    \item Parameters $\beta_i$s represent the exposure of stock $i$ to the market factor ${\rm d}f$;
    \item The time series $({\rm d}\epsilon^{i}_{t})_i$ represent the residuals of each stock returns, i.e. the portion of the stock returns not accounted for by market movements. These residuals are uncorrelated with the market;
    \item $\Omega_t$ is the log-volatility  stochastic \textit{factor} involved in the dynamics of both the market factor $(f_{t})_{t>0}$ {\it and} the residuals $(\epsilon^{i}_{t})_i$. We will choose to shift $\Omega_t$  such that the volatility of the market is equal to unity, i.e,  $\mathbb{E}[\exp(\Omega_t)]=1$,.
    \item $\omega^{i}_t$ is the stochastic \textit{idiosyncratic} log-volatility of residual returns ${\rm d}\epsilon^{i}$. Without loss of generality, we consider the mean of $\omega^{i}_t$ such that the volatility of the residual is directly $\sigma_i$, i.e, $\mathbb{E}[\exp(\gamma_i\Omega_t+\omega^{i}_t)]=1$.
    \item Parameters $\gamma_i$s represent the exposure of  the log-volatility of stock $i$ to the log-volatility of the index. The $\gamma$s are thus equivalent, in volatility space, to the usual factor model $\beta$s for returns.
    \item Finally, $W^0_t$ and $B^{i}_t$ are standard Brownian motions, all independent with one another.
\end{itemize}    


Drawing from this approach, we build in the next Section a Nested Log S-fBM one-factor model in which the market volatility and the idioyncratic volatilities are all Log S-fBM. 
\subsection{Definition of the N-SfFM: A Nested Factor Model with S-fBM log-volatilities}\label{sec:model}
We introduce the Nested factor model incorporating S-fBM processes as log-volatilities. For the sake of tractability and parcimony, we limit ourselves to the one-factor case, $M=1$, but our framework can be readily generalized to $M>1$ (see Appendix~\ref{app:MfactorNFM}). The idea is simply to complement the definition of the 1-factor NFM, Eq. \eqref{eq:NFM_definition} above, with the following extra specifications:
\begin{itemize}
    \item $\Omega_t$ is a S-fBM process of intermittency $\lambda^2$, Hurst exponent $H \in (0,1/2)$ and decorrelation parameter $T>0$;
    \item For any $i\in \llbracket 1, N\rrbracket$, $\omega^{i}_t$ is a S-fBM process of intermittency $\lambda_i^2$, Hurst exponent $H_i\in (0,1/2)$ and the same decorrelation parameter $T$.
\end{itemize}
We further assume that the market volatility mode $\Omega_t$ and the idiosyncratic modes $\left(\omega^{i}_t\right)_{i\in \llbracket 1,N \rrbracket}$ are all mutually independent. Furthermore, we suppose that the volatility modes are independent from the driving Brownian motions enabling no leverage effect which will be the object of future works.
The Nested Log S-fBM factor model (N-SfFM) is thus defined similarly to the original  version of Chicheportiche \textit{et al.} in \cite{chicheportiche2015nested};
the specificity of the present version is that we consider a S-fBM dynamic for the log-volatility modes in order to encode the roughness of their sample paths. Note that the correlation of the factor volatility mode and the residual log-volatility $\tilde{\omega}^{i}$ is parameterized by the multiplicative terms $\left(\gamma_i\right)_{i\in \llbracket 1,N\rrbracket}$ as we have:
\begin{eqnarray}
\label{eq:covfactorresidualmodes}
    \forall i \in \llbracket 1,N \rrbracket,\quad\forall t \geq 0, \quad\text{Cov}\left(\Omega_t,\tilde{\omega}^{i}_t \right)=\gamma_i \; \text{Var} \left(\Omega_t \right), \quad 
\end{eqnarray}

As evoked in Section \ref{sec:logSfBM}, we will leverage the small intermittency approximation, introduced in \cite{wu2022rough} initially used in the Log S-fBM calibration procedure, in the context of the log N-SfFM. Using the small intermittency approximation tool, we will show that under mild conditions on the exposure vectors $\left(\beta_i\right)_{i\in \llbracket 1,N \rrbracket}$ and $\left(\gamma_i\right)_{i\in \llbracket 1,N \rrbracket}$, observing the log-volatility of single stocks is equivalent to observing the log-volatility of the residuals, and observing the log-volatility of the index is equivalent to observing the log-volatility of the market factor. In other words, the residuals reflect the roughness of the single stock and the factor the roughness of the index.

\section{The small intermittency approximations}
\label{sec:SIA}
The small intermittency approximation is a tool initiated by Bacry\textit{ et al.} in \cite{bacry2008log} in order to study the scaling properties of the multifractal random walk model (see \cite{bacry2001multifractal} and \cite{muzy2000modelling}) and also used by Wu\textit{ et al.} in \cite{wu2022rough}. It consists in deriving an approximation of the generalized moments of the multifractal random measures in terms of the generalized moments of integrated processes of the type of Eq.~\eqref{eq:integratedlogvol} when the dimensionless intermittency parameter $\lambda^2$,  defined in Eq. \eqref{eq:rel_nu_lambda2}, goes to 0. Given that empirical values of $\lambda^2$ estimated from market data (see e.g. \cite{wu2022rough}) are around $\lambda^2 \simeq 0.05$, this approximation is shown to be accurate (more details are provided in Appendix \ref{app:accuracysmallintermittencyapproximationconsistency}).  The small intermittency approximation is relevant because none of the underlying S-fBM processes in Eq.~\eqref{eq:NFM_definition} are directly observable, as was also the case in \cite{wu2022rough}. Indeed, Wu\textit{ et al.} derived a small intermittency approximation of the autocovariance function of the log-multifractal random measure as well as the log-quadratic variation of a Log S-fBM process -- see Proposition 5 in \cite{wu2022rough}.\\
Here, we leverage the small intermittency approximation to derive a closed-form expression of the variance of the log volatility increments of a sum of multiple Log S-fBM processes, from which a similar result is derived in the case of our model in Eq.~\eqref{eq:NFM_definition} and that will be used extensively in the following Section.\\\\
For this sake, we consider $d$ Log S-fBM models:
    \begin{equation}
    \label{def:X_i}
    \forall\; i\in \llbracket 1,d \rrbracket,\;
        \begin{cases*}
      {\rm d}Y^{i}_t = e^{\frac{\omega^{i}_t}{2}} {\rm d}B^{i}_t\\
      Y^{i}_0=y^{i}_0 
    \end{cases*},
    \end{equation}
where $\left(\omega^{i}_t,i\in \llbracket 1,d \rrbracket\right)$ are $d$ independent S-fBM processes, each of which with respective parameters $\lambda_i,H_i,T$, $\left(B^{i},i\in \llbracket 1,d \rrbracket\right)$ is a $d$-dimensional Brownian motion (of independent entries) independent from $(\omega^{i}_t, i\in \llbracket 1,d \rrbracket)$ and $y^{i}_0$, $i=1, \ldots,d$ are constant values.

In what follows, the intermittency vector will be denoted as $\LL$ :
\begin{eqnarray}
\label{def:lambda_vec}
\LL = \begin{pmatrix}
\lambda_1 \\
\lambda_2\\
\vdots \\
\lambda_d
\end{pmatrix}
\end{eqnarray}

We introduce the weighted averaged S-fBM process $\omega_t$ and the weighted averaged process $\left(Y_t\right)_t$ 
as follows:
\begin{eqnarray}
\label{def:omega_X}
\begin{cases}
    \omega(t)=\sum_{i=1}^d a_i \omega^{i}_t\\
    Y_t:=\sum_{i=1}^d b_i Y^{i}_t
\end{cases},
\end{eqnarray}
 where $\left(a_i \right)_{i\in \llbracket 1,d \rrbracket}$ and $\left(b_i \right)_{i \in \llbracket 1,d \rrbracket} $ are tuples from $\R^{*d}$.\\\\
Along the same line as definition \eqref{eq:NFM_MRM_M_def}, for an arbitrary interval $K\subset \R_+$ such that $|K|\leq T$, the volatility measure built from $\exp(\omega_t)$ is denoted as $M$,
i.e, 
\begin{eqnarray}
\label{eq:MRMaggregexp}
    \forall I\subset K,\tab  M(I):=\int_{I}e^{\omega(t)}dt \; .
\end{eqnarray}

We introduce as well the quadratic variation of $Y$ of Eq.~\eqref{def:omega_X} obtained as the quadratic weighted sum of $Y^{i}$'s volatility measures:
\begin{eqnarray}
\label{eq:MRMaggreg}
    \forall I\subset K,\tab  \Sigma(I):=\sum_{i=1}^d b_i^2 M^{i}(I),
\end{eqnarray}
where $\left(M^{i}(I)  = \int_I e^{\omega^{i}_u} du \right)_{i\in \llbracket 1,d \rrbracket}$.\\
We introduce the respective volatility measures over a period of length $\Delta$:
\begin{eqnarray}
\label{eq:SigmaMdelta}
\forall t \geq 0,\tab
\begin{cases}
M_{\Delta}(t):=M([t,t+\Delta])\\
\Sigma_{\Delta}(t):=\Sigma([t,t+\Delta])
\end{cases}  .
\end{eqnarray}
For any $\tau>0$, we define  $W\left(\tau,\Delta,\LL\right)$ as the variance of $\ln\left( M_{\Delta}(\tau )\right) - \ln\left( M_{\Delta}(0)\right)$ representing the log-volatility increment of the volatility measure of Eq.~\eqref{eq:MRMaggregexp} during a period of length $\Delta$:
\begin{equation}
\label{eq:defW}
 W(\tau,\Delta,\LL) = \text{Var} \Big[ \ln\left( M_{\Delta}(\tau )\right) - \ln\left( M_{\Delta}(0)\right) \Big].
\end{equation}
In the same way, the variance of the random variable $\ln\left( \Sigma_{\Delta}(\tau )\right) - \ln\left( \Sigma_{\Delta}(0)\right)$, which represents the log volatility increment of the aggregated process $X$ during a period of length $\Delta$, is a function of the parameters $\tau$, $\Delta$ and $\LL$. So, we will denote it $V\left(\tau,\Delta,\LL\right)$:
\begin{equation}
\label{eq:defV}
 V(\tau,\Delta,\LL) = \text{Var} \Big[ \ln\left( \Sigma_{\Delta}(\tau )\right) - \ln\left( \Sigma_{\Delta}(0)\right) \Big].
\end{equation}

In what follows we will consider the limit of small intermittency 
and for that purpose we denote as respectively  $\mathcal{V}\left(\tau,\Delta,\LL \right)$ and 
$\mathcal{W}\left(\tau,\Delta, \LL \right)$ the first order approximation of $V(\tau,\Delta, \LL )$ and 
$W(\tau,\Delta,\LL)$, i.e.,:
\begin{eqnarray}
\label{eq:variancesumsmallintermiteq}
W\left(\tau,\Delta,\LL \right) & = & \mathcal{W}\left(\tau,\Delta,\LL\right)+o\left(\|\LL \| ^2\right) \\
V\left(\tau,\Delta,\LL \right) & = & \mathcal{V}\left(\tau,\Delta,\LL\right)+o\left(\|\LL \| ^2\right) \; 
. 
\end{eqnarray}
In Appendix \ref{app:proofoftheovariancesumsmallintermit}, we prove the following
result:
\begin{theo}
\label{theo:variancesumsmallintermit}
For any $(\tau,\Delta)\in K$, the small intermittency variances can be expressed as follows:
\begin{eqnarray}
\label{eq:variancesumsmallintermitequationexp}
 &\text{(a)} \;\;\;\;   \mathcal{W}\left(\tau,\Delta,\LL\right) =& \sum_{l=1}^d  a_l^2 \lambda_{l}^2\left(\frac{\tau}{T}\right)^{2H_l}g_{H_l}\left(\frac{\Delta}{\tau}\right). \\
 &\text{(b)} \;\;\;\;    \mathcal{V}\left(\tau,\Delta,\LL \right) = & \sum_{l=1}^d  b_l^4 \lambda_{l}^2\left(\frac{\tau}{T}\right)^{2H_l}g_{H_l}\left(\frac{\Delta}{\tau}\right). \label{eq:variancesumsmallintermitequation}
\end{eqnarray}
where  for any $H\in]0,\frac{1}{2}[$:
\begin{equation}
 \label{eq:defg}
    g_{H}\left(z\right):=\frac{|1+z|^{2H+2}-2|z|^{2H+2}+|1-z|^{2H+2}-2}{z^2H(1-(2H)^2)(2H+2)}  \; .
\end{equation}
\end{theo}

\noindent
Let us notice that we have, for any $0 < z < 1$, when $H \to 0$:
\begin{equation}
\label{eq:devg1}
g_{H} (z) = \frac{(z + 1)^2 \ln(z + 1) + (-1 + z)^2 \ln(1 - z) - 2z^2 \ln(z)}{z^2} + O(H)
\end{equation}
while, for any $0<H<1/2$, when $z \to 0$:
\begin{equation}
\label{eq:devg2}
g_{H}(z) = \frac{1}{H(1-2H)} + O(z^{2H})
\end{equation}
meaning that if we additionally have $H \to 0$, the equivalence holds:
\begin{equation}
g_{H}(z) \sim \frac{1}{H}
\end{equation}
\section{Dominant Hurst Exponents \&  Conditions on Model Parameters}\label{sec:conditions_on_parameters}

In order to account for the empirical findings of \cite{wu2022rough} within the N-SfBM, one should be in a situation where the factor log-volatility has a larger Hurst exponent (i.e. $H \sim 0.11$) than the log-volatility of single stocks that is observed to be close to $H_i = 0$ (see Figure \ref{fig:wu_empirical_results}). Now, as per Eq.~\eqref{eq:NFM_definition}, the stock returns are explained by the market factor contribution as well as the residual contribution. Hence the roughness of the single stocks should predominantly be explained by the roughness of the residuals. 

In the light of Ref. \cite{wu2022rough}, this suggests that the Hurst exponent of the factor's log-volatility should be higher than the one of the residual log-volatilities $\left(\tilde{\omega}^{i}\right)_{i\in \llbracket 1,N\rrbracket}$ ($\tilde{\omega}^{i} := \gamma_i \Omega + \omega^{i}$). In the following, we will derive conditions on $\left(\gamma_i\right)_{i\in \llbracket 1,N \rrbracket}$ as well as on the exposure vector $\left(\beta_i\right)_{i\in \llbracket 1,N \rrbracket}$ such that this is the case. All arguments developed below are based on estimation of the Hurst parameter based on the small intermittency approximation within the identification approach detailed in Appendix  \ref{subsubsec:hurstindentification}.

\subsection{Hurst exponent of the residual volatility -- Condition on $\gamma_i$}
\label{subSection:Hurstexponentofthe residualvolatilitymode}

For any $i\in \llbracket 1,N\rrbracket$, $\Omega$ and $\omega^{i}$ being (independent) Gaussian processes, $\tilde{\omega}^{i}$ is a Gaussian process too.\\
We consider the volatility measure associated with the process $\tilde{\omega}^{i}$ for a given $i\in \llbracket 1,N \rrbracket$:
\begin{eqnarray}
    \forall I \subset K,\tab  \widetilde{M}^{i}\left(I\right):=\int_Ie^{\tilde{\omega}^{i}(t)}dt
\end{eqnarray}
and for any $(t,\Delta)\in \R_+^{2}$, we denote similarly to Eq.~\eqref{eq:mrmdelta} the variance measure over a period of length $\Delta$:
\begin{eqnarray}
    \forall t\geq 0, \tab \widetilde{M}^{i}_{\Delta}(t):= \widetilde{M}^{i}\left([t,t+\Delta]\right).
\end{eqnarray}
Indeed, $\tilde{\omega}^{i}$ is a linear combination of two independent S-fBM processes for any $i\in \llbracket 1,N \rrbracket$. Thanks to the point (a) of Theorem ~\ref{theo:variancesumsmallintermit}, $\widetilde{\mathcal{W}}_i\left(\tau,\Delta,\LL_i\right)$, the small intermittency approximation of the variance of the log-volatility increments of the residuals during a period of length $\Delta$ denoted $\ln\left( \widetilde{M}^{i}_{\Delta}(\tau )\right) - \ln\left( \widetilde{M}^{i}_{\Delta}(0)\right)$ (see the definition in  Eq.~\eqref{eq:variancesumsmallintermiteq}) can be expressed as follows:

\begin{eqnarray} \label{eq:gamma_iconditionvariance}
       \widetilde{\mathcal{W}}_i\left(\tau,\Delta,\LL_i\right)= \gamma_i^2 \lambda^2\left(\frac{\tau}{T}\right)^{2H}g_{H}\left(\frac{\Delta}{\tau}\right)+ \lambda_{i}^2\left(\frac{\tau}{T}\right)^{2H_i}g_{H_i}\left(\frac{\Delta}{\tau}\right), \nonumber
\end{eqnarray}
where:
\begin{itemize}
\item $\LL_i$ is the intermittency vector $\begin{pmatrix} \lambda \\ \lambda_i  \end{pmatrix}$
    \item $g_H(z)$ defined in Eq. \eqref{eq:defg}
\end{itemize}

According to the approach advocated in Appendix \ref{subsubsec:hurstindentification},
if the $\gamma_i^2 \lambda^2\left(\frac{\tau}{T}\right)^{2H}g_{H}\left(\frac{\Delta}{\tau}\right)$ contribution to the variance of $\ln\left( \Tilde{M}^{i}_{\Delta}(\tau )\right) - \ln\left( \Tilde{M}^{i}_{\Delta}(0)\right)$ was dominant, the residual volatility  would have the Hurst exponent as that of the market factor, which contradicts the findings of \cite{wu2022rough}. As a result, $\gamma_i$ should be such that:\footnote{Note: throughout the following heuristic discussion, $a\ll b$ means that $b$ is more than -- say -- 10 times $a$.}
\begin{equation}
\label{ineq1}
 \gamma_i^2 \ll \frac{\lambda_i^2}{\lambda^2}  \left(\frac{\tau}{T}\right)^{2H_i-2H} r_{H_i,H}\left(\frac{\Delta}{\tau} \right)
\end{equation}
where we have denoted 
\begin{equation}
\label{eq:def_r}
   r_{H_i,H}(z) = \frac{g_{H_i}\left(z\right)}{g_{H}\left(z\right)} \; .
\end{equation}

If we refer to the estimates in \cite{wu2022rough}, we can furthermore assume that  $\lambda_i \simeq \lambda$,  $H \simeq 0.1 \gg H_i \simeq 0$. In this case, using expression \eqref{eq:defg}, that, one can show that, $\forall z < 1$ $r_{0,H} \geq 1$. It results that a sufficient for  
condition \eqref{ineq1} to hold is that:
$$
   \gamma_i \ll \left( \frac{T}{\tau}\right)^{H}   \; .
$$
which holds, in the range $\tau \ll T$, provided
\begin{equation}
\label{eq:gamma_criterion}
 \gamma_i \leq 1 \; .
\end{equation}

From the empirical results of \cite{chicheportiche2015nested}, we know that $\gamma_i < 0.5$ for all stocks so that, according to previous criterion, the residual volatility roughness is dominated by $H_i$ for any $\tau \ll T$.

\subsection{Hurst exponents of single stocks -- Upper bound on $\beta_i$}
\label{subSection:Hurstexponentsofsinglestocksandstockindexes}

Let us now exhibit an upper bound condition on the exposure coefficient $\beta_i$ in order for the Hurst exponent of the idiosyncratic residual volatility mode to represent the one of stock $i$. Again, our discussion will rely on the identification approach detailed in  Appendix~\ref{subsubsec:hurstindentification}.

Indeed, any ${\rm d}x^{i}$ is in the same configuration as in the Section \ref{sec:SIA}. Under the condition \eqref{eq:gamma_criterion}, we can therefore use 
point (b) of Theorem~\ref{theo:variancesumsmallintermit}
to obtain the following expression of $\mathcal{V}_i\left(\tau,\Delta,\LL_i\right) $, the small intermittency approximation of the variance of $\ln\left( M^{i}_{\Delta}(\tau )\right) - \ln\left( M^{i}_{\Delta}(0)\right)$ the log-quadratic variation increment of $x^{i}$ during a period of length $\Delta$:
\begin{eqnarray} \label{eq:beta_contribution}
        \mathcal{V}_i\left(\tau,\Delta,\LL_i\right) = \beta_i^4 \lambda^2\left(\frac{\tau}{T}\right)^{2H}g_{H}\left(\frac{\Delta}{\tau}\right)+\sigma_i^4 \lambda_i^2\left(\frac{\tau}{T}\right)^{2H_i}g_{H_i}\left(\frac{\Delta}{\tau}\right) \; . \nonumber
\end{eqnarray}

Using the same arguments of the one in Section ~\ref{subSection:Hurstexponentofthe residualvolatilitymode}, the idiosyncratic volatility mode is dominated by the second term of Eq. \eqref{eq:beta_contribution} provided: 
\begin{equation}
\label{eq:ineq2}
\beta_i^4  \ll \frac{\sigma_i^4 \lambda_i^2}{\lambda^2}  \left(\frac{\tau}{T}\right)^{2H_i-2H} r_{H_i,H}\left(\frac{\Delta}{\tau} \right)
\end{equation}
where $r_{H_i,H}(z)$ is defined in \eqref{eq:def_r}. Along the same line as for Eq. \eqref{ineq1},
given the empirical values of $H,H_i,\lambda,\lambda_i$, \eqref{eq:ineq2} holds in the range $\tau \ll T$
if:
\begin{equation}
\label{eq:beta_criterion}
| \beta_i |  \leq \sigma_i 
\end{equation}

This condition appears to be empirically valid since 
the idiosyncratic contribution to volatility $\sigma_i$ is around $2$ while factor exposure $\beta_i$ appear to be in the range $(0.5 - 1.5)$ (see e.g. \cite{garnier2021new} as well as Appendix \ref{app:miscnumerics} in particular Figure~\ref{fig:distributionbetas}, \ref{fig:distributionsigmas} and \ref{fig:distributionbetasoversigmas}), with some exceptions: high-beta stocks with low idiosyncratic volatilities will presumably look more like the index, as indeed expected.

\subsection{Hurst exponent of sub-indexes}

Let us consider sub-indexes $I_t$, constructed from a subset of size $N_s$ of stocks $x^{i}_t$ with weights $\left(w_i\right)_{i\in \llbracket 1,N_s \rrbracket} \in \R^{* N_s}$:
\begin{eqnarray}
\label{eq:index}
    I_t = \sum_{i=1}^{N_s} w_ix^{i}_t, \qquad \sum_{i=1}^{N_s} w_i=1.
\end{eqnarray}
Note that when ${N_s}=N$, the total number of stocks in an index, $I_t$ is simply the market mode $f_t$.
The sub-index evolution is given by:
\begin{eqnarray}
\label{eq:sub_index}
    {\rm d}I_t = \overline{\beta} {\rm d}f_t+\sum_{i=1}^{N_s} w_i{\rm d}e^{i}_t,
\end{eqnarray}
where we have denoted $\overline{\beta}$ the mean value of $\beta_i$:
\begin{equation}
    \overline{\beta}:= \sum_{i=1}^{N_s} w_i\beta_i.
\end{equation}

In Eq. \eqref{eq:sub_index} we are in the same configuration as in Section \ref{sec:SIA}. Conducting similar computations as previously, under individual condition \eqref{eq:gamma_criterion}, the small intermittency approximation of the variance of $\ln\left( M^{I}_{\Delta}(\tau )\right) - \ln\left( M^{I}_{\Delta}(0)\right)$ (the log-quadratic variation of the sub-index process $I$ during a period of length $\Delta$) reads:
\begin{eqnarray}
 \mathcal{V} \left(\tau,\Delta,\LL\right)=\overline{\beta}^4\lambda^2\left(\frac{\tau}{T}\right)^{2H}g_{H}\left(\frac{\Delta}{\tau}\right)+ \sum_{i=1}^{N_s}\ w_i^4 \sigma_i^4 \lambda_i^2\left(\frac{\tau}{T}\right)^{2H_i}g_{H_i}\left(\frac{\Delta}{\tau}\right). 
\end{eqnarray}

Under the same approach as before, we can see that the Hurst exponent of the sub-index $I$ in Eq.~\eqref{eq:index} can be approximated with the Hurst exponent of the factor $H$ if the first term of the previous sum dominates, namely:
\begin{equation} \label{eq:beta_cond_0}
 \overline{\beta}^4 \gg  \sum_{i=1}^{N_s} \frac{w_i^4 \sigma_i^4 \lambda_i^2}{\lambda^2} 
 \left(\frac{\tau}{T}\right)^{2(H_i-H)}  r_{H_i,H}\left(\frac{\Delta}{\tau}\right). 
\end{equation}
In the case when $H_i \to 0$ (as suggested by the empirical findings of \cite{wu2022rough}), a, $H_i \to 0$) one can show that (the proof is in Appendix \ref{app:proofineqr_0H}):
\begin{equation} \label{eq:ineqr_0H}
 \forall z \in ]0,1[,\tab r_{0,H}(z) \leq C_H\left(3-2 \ln(z)\right), 
\end{equation}
with $C_H = \frac{H(1-2H)(1+2H)(1+H)}{2\left(2^{2H}-1\right)}$. In that situation, provided $\lambda_i \simeq \lambda$, previous inequality holds if:
\begin{equation} \label{eq:beta_cond}
 \overline{\beta}^4 \gg   C_H \left(\frac{T}{\tau}\right)^{2H} \! \! \ln\left(\frac{e^3 \tau^2}{\Delta^2}\right) \overline{w^3 \sigma^4},
\end{equation}
where 
$$\overline{w^3 \sigma^4} = \sum_{i=1}^{N_s} w_i^4 \sigma_i^4 \; $$ 
represents the average value of $w_i^3 \sigma_i^4$, weighted by $w_i$. 
For the sake of simplicity, one can consider that $\sigma_i \simeq 2$ and $w_i = 1/N_s$ which gives 
$$\overline{w^3 \sigma^4} = \frac{16}{N_s^3}$$ leading to the condition:
\begin{equation} 
\label{eq:beta_cond}
 \overline{\beta}^{\frac{4}{3}} N_s  \gg   D_H^{\frac{1}{3}} \left(\frac{T}{\tau}\right)^{\frac{2H}{3}} \! \! \left( \ln\left(\frac{e^3 \tau^2}{\Delta^2}\right)\right)^{\frac{1}{3}}  
\end{equation}
with $D_H = 16 C_H$.

Given that $\Delta < \tau < T$, for $\Delta = 1$ trading day and $T \approx 5. 10^3 \; \Delta$ corresponding roughly an overall sample size $T$ of 10 years, with $H = 0.1$, one has:
\begin{equation*}
 \left(\frac{T}{\Delta}\right)^{\frac{2H}{3}} \! \! \left( \ln\left(\frac{e^3 T^2}{\Delta^2}\right)\right)^{\frac{1}{3}} \simeq  5 \tab \text{and} \tab 
 D_H^{1/3} \simeq 1.8 \; .
\end{equation*}
It results that condition \eqref{eq:beta_cond} holds provided
\begin{equation} 
\label{eq:beta_cond_final}
 \overline{\beta}^{\frac{4}{3}} N_s \gg 10 \; .
\end{equation}
The latter condition can be understood from two perspectives: for relatively small $\beta$-exposures, one needs to have a large basket of single stocks to reach the regime where the regularity of the index is encoded in the regularity of the factor's sample paths. If on the contrary the basket of single stocks is of relatively small size $N_s$, those single stocks would better have a high exposure to the factor so that the factor represents well the overall index dynamics (corresponding to $N_s=N$).  \\

That being said, for a fixed $N_s$, and given condition Eq.~\eqref{eq:beta_criterion} from the previous Section,
a sufficient condition for any $i\in \llbracket 1,N_s \rrbracket$ on each exposure value summarized as
\begin{eqnarray}
\label{eq:conditiononbetas}
  \left( \frac{10}{N_s} \right)^{\frac{3}{4}} \ll \left| \beta_i \right|
\ll \sigma_i.
\end{eqnarray}

In the next Sections, we give some results concerning the empirical estimation of the N-SfFM, then present numerical experiments confirming the theoretical estimates made in this Section. We will then turn to an empirical validation of our scenario.

\section{Calibration Method}
\label{sec:calibrationmethod}
This section describes the calibration procedure we use in order to estimate respectively  the Hurst exponent of the factor $H$ and the Hurst exponents of the idiosyncratic modes $H_i$. We establish an asymptotic formula, when the number of single stocks $N_s$ goes to infinity, of the estimated factor quadratic variation in terms of the assets quadratic variations and the exposure vector $\beta$ that is necessary to estimate its Hurst exponent via the general method of moments (GMM) (see \cite{bolko2020roughness}) used by Wu\textit{ et al.}  (in practice, we consider moments of second order: autocovariance functions, following the approach described in Section 4 in \cite{wu2022rough}), which is applied in Section \ref{Section:numerics}.\\

Our calibration method is decomposed in the following steps:
\begin{enumerate}[label=\textbf{Step~\arabic*}, wide, labelwidth=!, labelindent=0pt]
\setlength\itemsep{0.1em}
    \item consists in estimating the exposure vector $\beta$. Utilizing the estimation approach proposed by Chicheportiche\textit{ et al.} in \cite{chicheportiche2015nested}, we aim to determine $\beta$ so that $\beta\beta^\top$ serves as an approximation to the covariance matrix of the asset returns $({\rm d}x^{i}_t)_{i\in\llbracket1,N_s\rrbracket}$, represented by $XX^\top$:
\begin{eqnarray}
\label{eq:betaestimation}
    \hat{\beta}=\underset{\beta\in \R^{N_s}}\argmin && \left|\left|\E\left(XX^\top\right)-\beta\beta^\top\right|\right|_{\text{off diagonal}},
\end{eqnarray}
where $\left|\left| . \right|\right|_{\text{off diagonal}}$ is the off-diagonal Frobenius norm.\\
The use of the off-diagonal Frobenius norm was already proposed by Chicheportiche \textit{et al.} in \cite{chicheportiche2015nested} for calibrating the exposure vector $\beta$. Its use here is for calibrating the exposure vector focusing on the asset covariations.\\
Thanks to the formulation of the model in Eq.~\eqref{eq:NFM_definition}, the correlation between the residual volatility modes does not play any role since the residual driving Brownian motions are independent.\\
    \item aims at identifying a proxy for the factor quadratic variation, denoted as $\left<f\right>_t$. This component is crucial for estimating the factor Hurst exponent $H$. Our approach utilizes Eq.~\eqref{eq:qvfactor} as presented in the following result.

\begin{prop}
\label{prop:qvfactorintermsofqvassets}
We consider ${N_s}\in \N^{*}$. 
If $\beta$ has independent and identically distributed entries such that:
\begin{eqnarray}
\forall i\in \llbracket 1,{N_s} \rrbracket, \tab \E\left(\beta_i^4\right) < \infty \nonumber    
\end{eqnarray}
and independent of $\left(\epsilon_i\right)_{i \in \llbracket 1,{N_s} \rrbracket}$, the following claim holds :
    \begin{eqnarray}
    \label{eq:qvfactor}
    \forall I\subset K,\quad\tab \left<\hat{f}\right>_I \underset{{N_s}\rightarrow\infty}\sim \frac{1}{\sum_{i=1}^{N_s}\beta_i^4}\sum_{i=1}^{N_s}\beta_{i}^2\left<x^{i}\right>_I, \tab a.s ,
\end{eqnarray}
where $\hat{f}$ is defined as:
\begin{eqnarray}
\label{eq:olsfactor}
\forall t\in K,\quad
 \widehat{f}_{t}:=\underset{f\in \R}\argmin  \left|\left| X_t-\beta f\right|\right|,
\end{eqnarray}
where $\left|\left| .\right|\right|$ is the $\mathcal{L}^2$ norm in $\R^{N_s}$.
\end{prop}
The proof is given in Appendix \ref{app:proofqvfactorintermsofqvassets}.\\\\

It is noteworthy that Eq.~\eqref{eq:qvfactor} allows for the direct estimation of the factor quadratic variation using the quadratic variations of asset returns. Consequently, in practical applications, the volatility of the factor can be directly inferred from the volatility of individual stocks by employing a daily volatility estimator based on high-frequency (intraday) data, such as the Garman-Klass estimator (see Eq.~\eqref{eq:GK-est}). This approach is contingent upon having a sufficiently large number of stocks in the sample, as illustrated in our empirical study in Section~\ref{Section:empirical}.\\\\

The log-quadratic variation of returns serves as the principal input for the Generalized Method of Moments (GMM) introduced in \cite{wu2022rough} and employed to estimate the roughness, specifically the Hurst exponent, of a given time series based on the small intermittency approximation of covariance of log volatilities. As explained in Appendix \ref{app:intermittencybias}, the bias on the estimated parameters is mitigated in such limit. The GMM operates under the assumption that log-volatility follows a Gaussian process. Therefore, it is prudent, prior to estimating roughness, to address the pronounced skewness apparent in the underlying time series. This skewness results from the logarithmic transformation, which tends to amplify the effects of lower return values. Such skewness subsequently propagates into the log-volatility (and, by extension, the log-quadratic variation) components, thereby contravening the Gaussian assumption inherent in the SfBM model and adversely impacting the estimation of the Hurst exponent, particularly when using daily data. To comply with this assumption, we undertake a \textit{``Gaussianisation''} procedure as a  pre-processing step  in order to rectify the empirically determined log-quadratic variations that are distorted by measurement noise. The applied methodology is detailed in  Appendix~\ref{app:Gaussian_trick}.\\\\

Hereafter, we consider as the estimated time series of $\left(\Omega_t\right)_t$ the gaussianized version of $\ln\left(\left<f\right>_t\right)-\E\left( \ln\left(\left<f\right>_t\right)\right)$.
    \item leverages the GMM introduced in \cite{wu2022rough} to estimate $H$ and $\lambda^2$ from the time series of $\left(\Omega_t\right)_t$ derived in the previous step. 

    \item deduces the estimated quadratic variations of the residual processes $\left(\left<\epsilon^{i}\right>_t\right)_{i\in\llbracket1,N_s\rrbracket}$ through the identity:
\begin{eqnarray}
\label{eq:qvresiduals}
    {\rm d}\left<x^{i}\right>_t = \beta_i^2 \, {\rm d}\left<f\right>_t + {\rm d}\left<\epsilon^{i}\right>_t.
\end{eqnarray}
     
    \item aims at estimating the Hurst exponents $H_i$ of the idiosyncratic log-volatility of individual stocks. The process involves the following steps:
    \begin{itemize}
        \item The time series $\left(\tilde{\omega}_t^{i}\right)_{i\in\llbracket1,N_s\rrbracket}$ are approximated using the gaussianized versions of $\ln\left(\left<\epsilon^{i}\right>_t\right)-\E\left( \ln\left(\left<\epsilon^{i}\right>_t\right)\right)$, as derived in Step 4;
        \item Eq.~\eqref{eq:NFM_definition} enables the estimation of the time series of the idiosyncratic modes, $\left(\omega_t^{i}\right)_{i\in\llbracket1,N_s\rrbracket}$, by regressing the time series $\left(\tilde{\omega}^{i}\right)_{i\in\llbracket1,N_s\rrbracket}$ against $\Omega_t$, derived in Step 1;
        \item Finally, the series $\left(\omega_t^{i}\right)_{i\in\llbracket1,N_s\rrbracket}$ serve as the input for the GMM to estimate the corresponding Hurst exponents $H_i$ as well as the parameters $\lambda_i^2$ for all $i\in\llbracket1,N_s\rrbracket$.
    \end{itemize}
\end{enumerate}

\section{Numerical Experiments}
\label{Section:numerics}

This Section is devoted to numerical experiments. First, we present numerical evidence that the N-SfFM  indeed reproduces the stylized facts discussed in Sections~\ref{subSection:Hurstexponentofthe residualvolatilitymode} and \ref{subSection:Hurstexponentsofsinglestocksandstockindexes}. More precisely, the Hurst exponent of the factor is expected to be the one of the index (constructed as in Eq.~\eqref{eq:index} with equal weights) and the Hurst exponent of the the single stocks is the one of the idiosyncratic residual modes. The formulation of Eq.~\eqref{eq:NFM_definition} allows a correlation between the volatility modes of the factor the residuals, which is one of the purposes of the Nested Factor Model of Chicheportiche\textit{ et al.} in \cite{chicheportiche2015nested}. Then, we demonstrate that the roughness of the index is indeed encoded in the roughness of the factor. Finally by using the calibration method detailed in Section~\ref{sec:calibrationmethod}, we evaluate how robust the estimate of the Hurst exponent $H$ of the factor as well as the Hurst exponents of the idiosyncratic modes $H_i$ are.\\

\begin{figure}[H]
    \centering
    \includegraphics[width=0.76\linewidth]{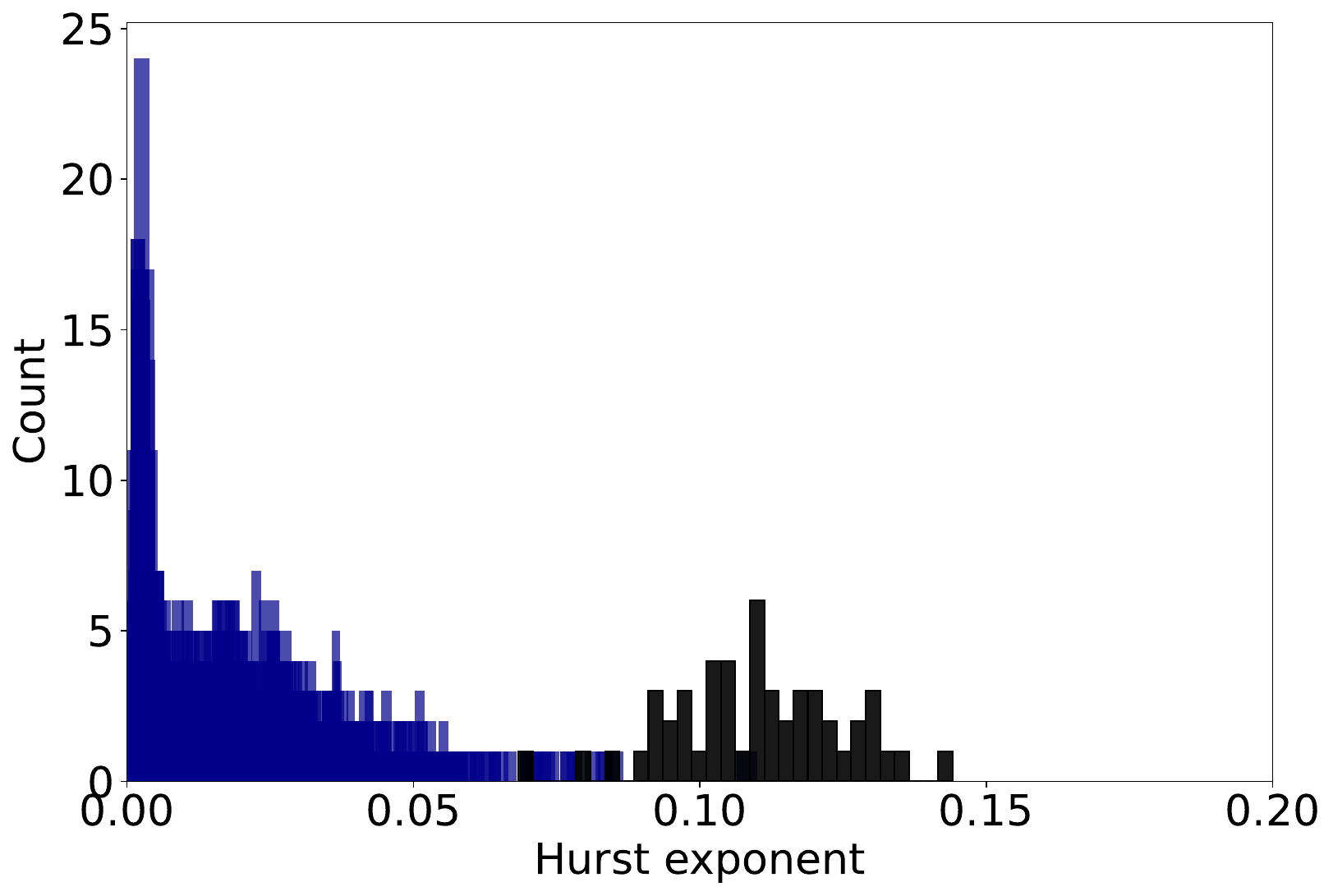} 
        \label{fig:subfig1}
    \caption{We construct the empirical distribution through a 50 sized sample of the estimated Hurst exponents of $N=100$ synthetic single stocks (\textbf{blue}) together with the one of the index (\textbf{black}) defined as the empirical mean of the simulated stocks. $\beta$ entries are considered independent and identically distributed  drawn from a  beta-distribution of parameters $a=9.6$ and $b=5.6$. The model parameters are fixed as follows: $H=0.11$, $\lambda=\lambda_i=0.05$,  $H_i=0.01$ and $\sigma_i=1$ for all $i\in \llbracket 1,N \rrbracket$, $L=2^{14}$, $T=2^{12}$, $s=2^8$ and $\gamma_i=0.2$ for all $i\in \llbracket 1,N \rrbracket$.}
    \label{fig:figstocksvsindex}
\end{figure}

In order to show that the N-SfBM reproduces the non-intuitive behaviour announced earlier, we generate the model with $N$ synthetic asset return trajectories $x_t^{i}$ of total
length $L$ following the dynamics specified by Eq.~\eqref{eq:NFM_definition}, as well as the index trajectories obtained as the average path over different $N$ assets with  $s=2^8$ simulated data points of the asset returns between the time $n\Delta$ and $(n+1)\Delta$ for $n=0,...,L-1$. 
The model parameters are specified accordingly.\\
To choose realistic values for $\beta_i$ and $\sigma_i$, one can use the empirical distributions of these coefficients from the asset returns of stocks that constitutes the S\&P500. This is documented in Appendix  \ref{app:betasigmaemiricaldistrib} where the vector $\beta$ is calibrated using close-to-close returns of an arbitrary bench of S\&P500 stocks using  Eq.~\eqref{eq:betaestimation} and the vector $\sigma$ is identified using the variance of the residuals obtained using the same bench of stocks. As a result, the distribution of the components of $\beta$ and $\sigma$ can be fitted as a beta-distribution whose density is of the form $x^{a-1} (1 - x)^{b-1}$. In order to respect the feasible conditions of Section \ref{sec:conditions_on_parameters}, we choose to fix $\sigma_i=1$ for all $i\in \llbracket 1,N\rrbracket$. \\\\
We present in Figure \ref{fig:figstocksvsindex} the empirical distributions of the Hurst exponents of the single stocks versus the empirical distribution of the index Hurst exponent generated as described earlier.

\begin{figure}[H]
    \centering
    \includegraphics[width=0.8\linewidth]{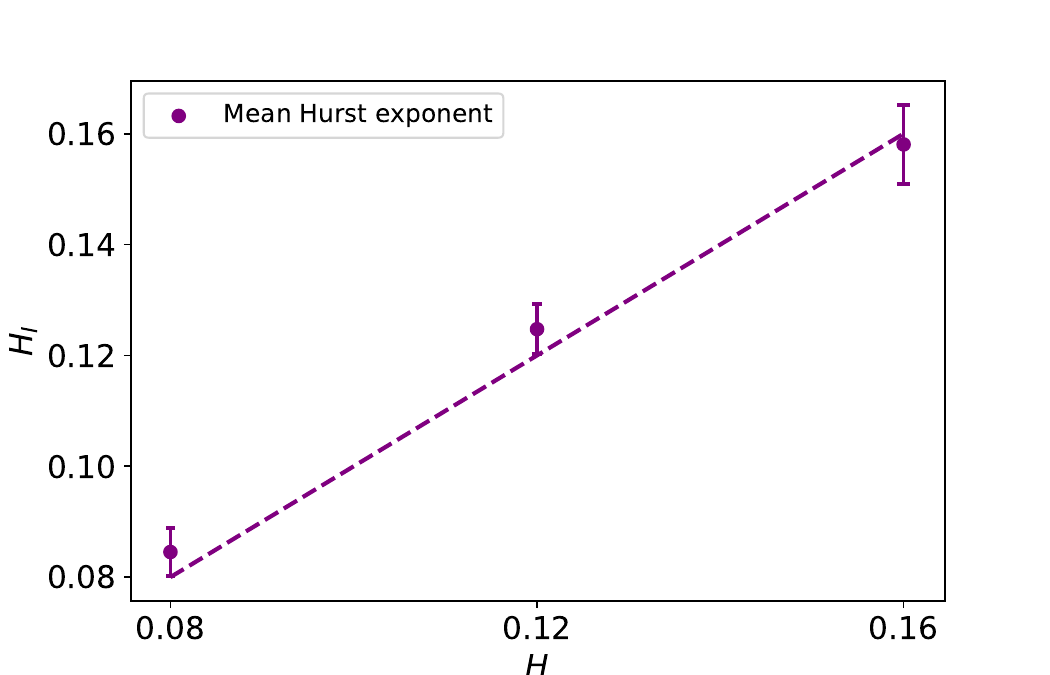} 
        \label{fig:subfig1}
    \caption{For each value of $H$, we construct a 20 sized sample of the estimated Hurst exponents of the empirical index $H_I$ constructed with $N=200$ synthetic single stocks and we plot its mean value together with the 95\% confidence interval (\textbf{purple}) as well as the dashed line that corresponds to the first bisector. $\beta$ entries are considered independent and identically distributed  drawn from a  beta-distribution of parameters $a=9.6$ and $b=5.6$. The model parameters are fixed as follows: $\lambda^2=\lambda_i^2=0.001$, $\sigma_i=1$,  $H_i=0.01$ for all $i\in \llbracket 1,N \rrbracket$, $T=L=2^{15}$, $s=2^8$ and $\gamma_i=0.01$ for all $i\in \llbracket 1,N \rrbracket$.}
    \label{fig:figHfactorvsHindex}
\end{figure}

We observe that the Hurst exponent distribution of the index is centered on the theoretical Hurst exponent $H=0.11$ of the factor, and the empirical distribution of the single assets Hurst exponents produces mainly values around $0.01$ which resembles Figure \ref{fig:wu_empirical_results}.\\\\
In Figure \ref{fig:figHfactorvsHindex} , we evaluate the discrepancy between the empirical distribution of the Hurst exponent of the index denoted $H_I$ and the Hurst exponent of the factor $H$. The latter index is constructed similarly as in the experiment of Figure \ref{fig:figstocksvsindex}. One can see that the estimated Hurst exponent of the index corresponds (within 95\% confidence interval) to the Hurst exponent of the factor. This means that the roughness of the index is captured by the roughness of the factor in the N-SfFM dynamic as expected from the condition of  Eq.~\eqref{eq:conditiononbetas}.

\begin{figure}[H]
    \centering
    \includegraphics[width=0.49\linewidth]{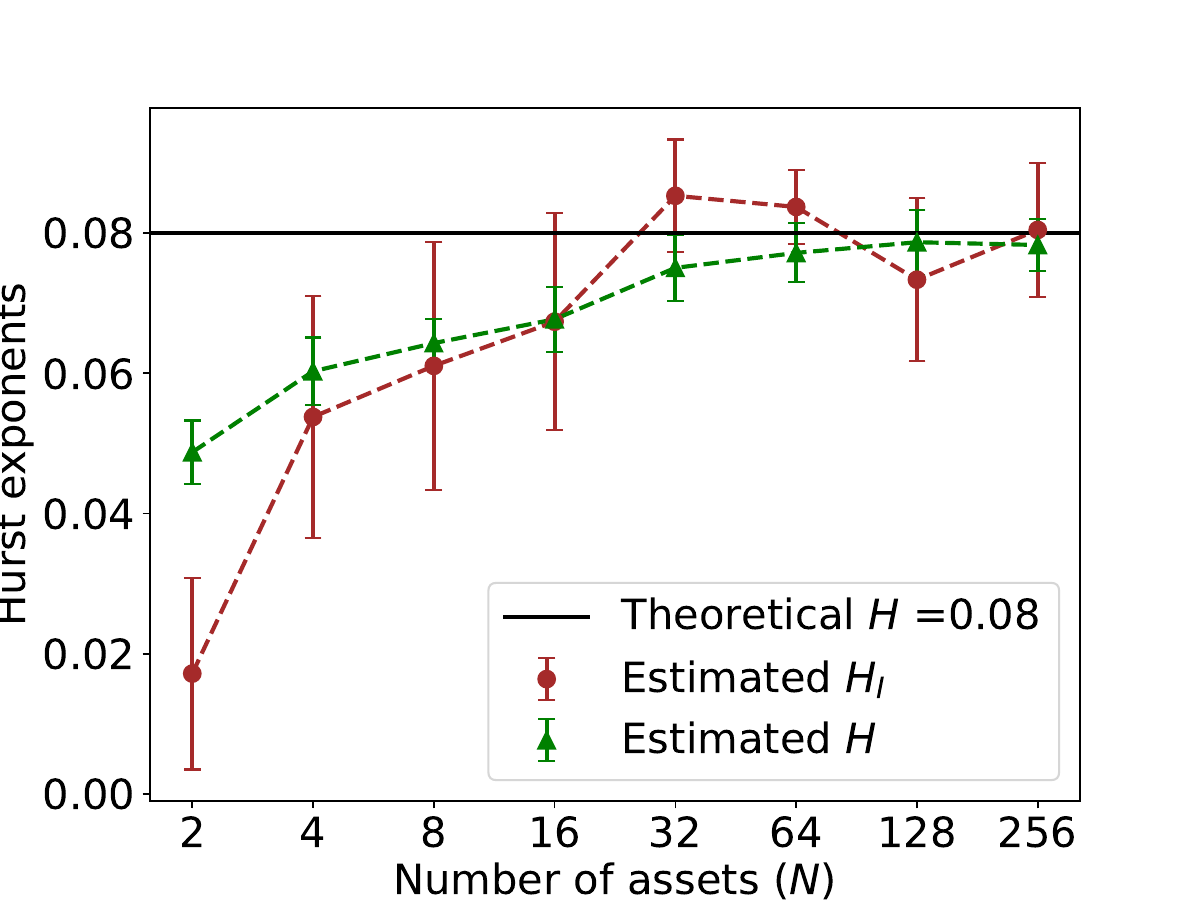} \hfill\includegraphics[width=0.49\linewidth]{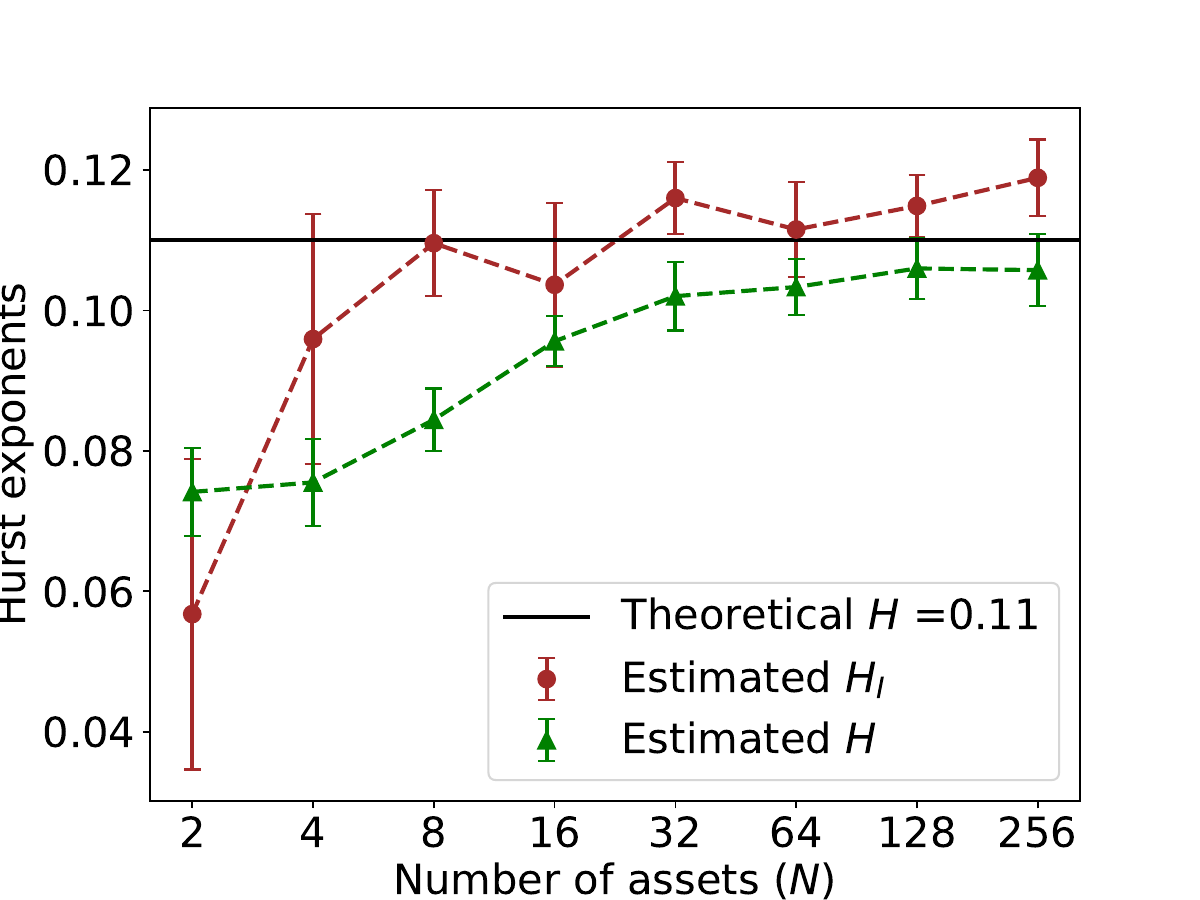}
    \caption{Using synthetic data for different index sizes $N$, we plot the mean value and the associated 95\% confidence interval from a sample of 20 copies of the Hurst exponent of the empirical index estimates in (\textbf{brown}) and the estimated Hurst exponent of the factor (\textbf{green}) together with error bars compared with the theoretical factor Hurst exponent  $H=0.08$ (\textbf{left}) and $H=0.11$ (\textbf{right}) (in horizontal \textbf{black} line). $\beta$ entries are considered to be independent and identically distributed  drawn from a beta-distribution of parameters $a=9.6$ and $b=5.6$, $\lambda^2=\lambda_i^2=0.001$, $\sigma_i=1$,  $H_i=0.01$ for all $i\in \llbracket 1,N \rrbracket$, $T=L=2^{15}$, $s=2^8$ and $\gamma_i=0.01$ for all $i\in \llbracket 1,N \rrbracket$.}
    \label{fig:figHurstindexandfactorVSN}

\end{figure}

In Figure~\ref{fig:figHurstindexandfactorVSN}, we evaluate the estimation bias of $H_I$ and $H$ through its empirical distribution and deviation around a chosen theoretical value of the factor Hurst exponent $H=0.08$ or $H=0.11$, as $N$ goes to infinity. For each value of $N$, we generate a sample of 20 batches of synthetic stock returns which enables to obtain a sample of 20 estimates of $H_I$ and $H$. Then we plot the associated mean values with the $95\%$ confidence interval in comparison with the theoretical value of $H$. As $N$ increases, we see that the estimated Hurst exponent of the index converges to the chosen theoretical values of the Hurst exponents of the factor. Furthermore, for small values of $N$ the index roughness is close to the one of a single stock as there is no sufficient number of assets to offset the effect of the residuals. As $N$ grows, one can see that estimated Hurst exponent of the index converges to the theoretical value of the factor Hurst exponent $H$. For small values of $H$ one can see that a higher number of assets is needed to reach the previous regime because of the exposure of single stocks to the factor. Similarly via applying the calibration algorithm, we see that, as $N$ grows, the estimated Hurst exponent of the factor also reaches the chosen theoretical values. Furthermore, we do observe a continuum of effective Hurst exponents for the dominant factor as $N$ increases: for small $N$ values, the dominant factor behaves as a single stock with $H\ll 0.11$ (see Proposition 2 in \cite{wu2022rough}); larger $N$ leads to a better defined factor with a Hurst exponent approaching its theoretical value. It is interesting to notice a small bias between the estimated Hurst exponent of the index and the factor on one side and between the estimated Hurst exponent of the factor and its corresponding theoretical value on the other side which may come from the error in proxying the quadratic variation of the factor (see step 2 of the calibration method in Section \ref{sec:calibrationmethod}). This will be the object of a future study.

\begin{figure}[H]
    \includegraphics[width=0.49\linewidth]{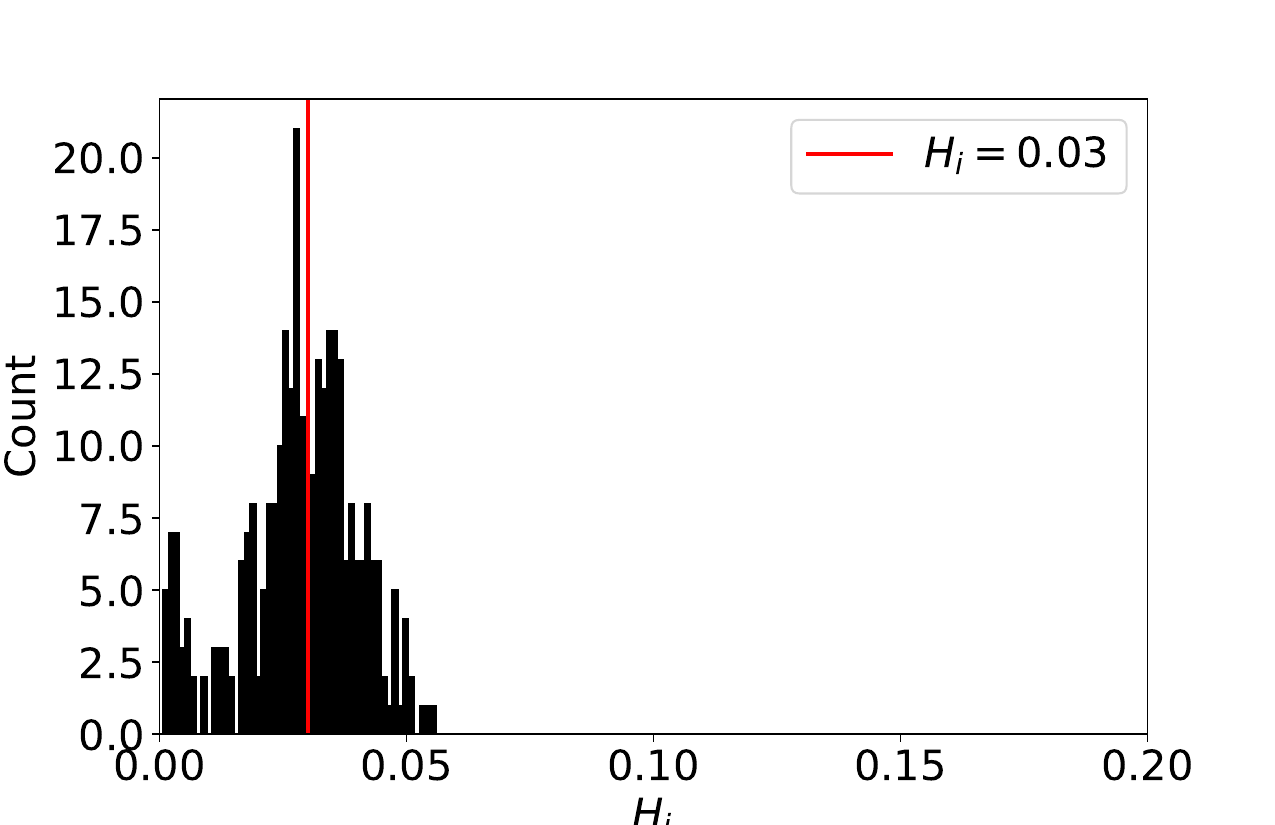}\hfill
    \includegraphics[width=0.49\linewidth]{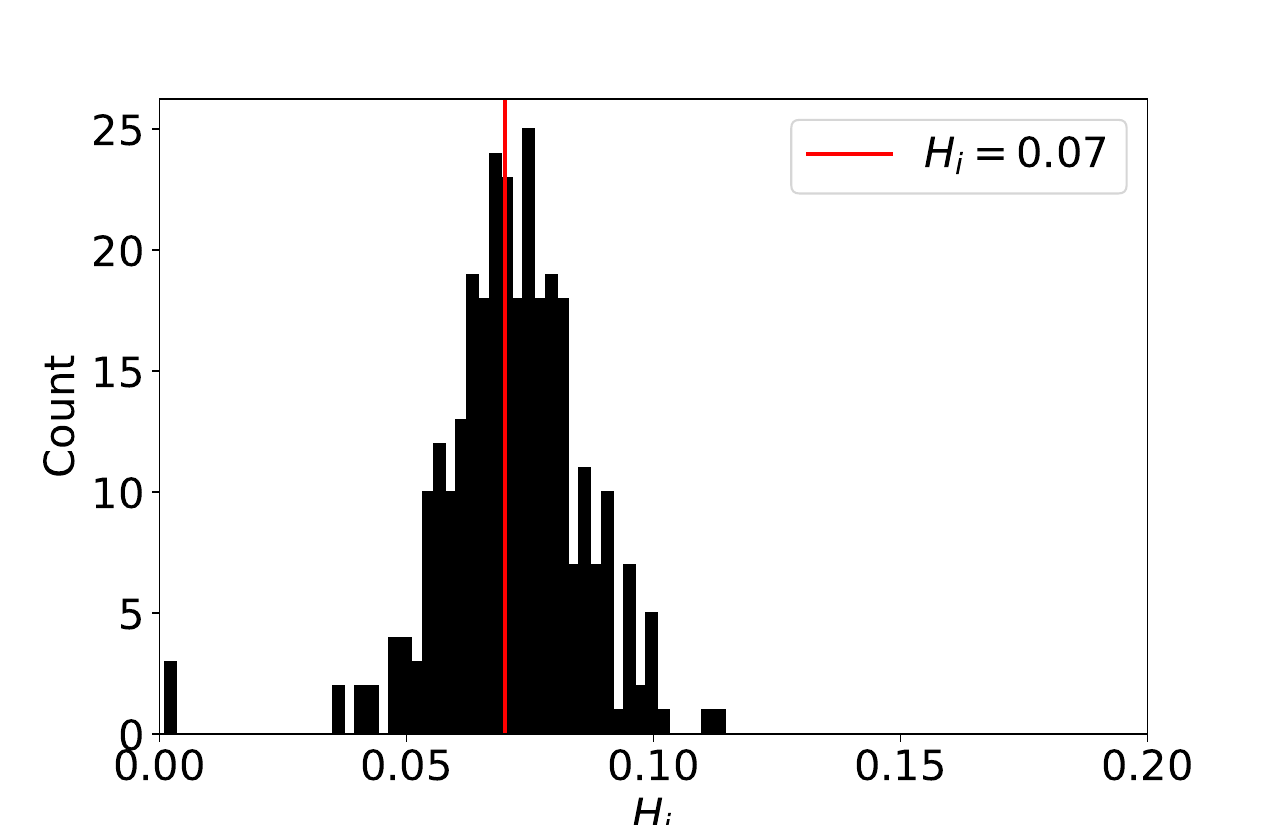}
    \caption{Calibration of the Hurst exponents $H_i$ using synthetically generated data for $N=300$ with the following parameters: $H=0.1$, $\beta$ entries are considered to be independent and identically distributed  drawn from a beta-distribution of parameters $a=9.6$ and $b=5.6$, $\lambda^2=\lambda_i^2=0.01$, $\sigma_i=1$,  $H_i=0.03$ (\textbf{Left}) and $H_i=0.07$ (\textbf{Right}) for all $i\in \llbracket 1,N \rrbracket$, $T=L=2^{15}$, $s=2^8$ and $\gamma_i=0.1$ for all $i\in \llbracket 1,N \rrbracket$.} 
    \label{fig:histogramsHi}
\end{figure}

Finally, we evaluate in Figure \ref{fig:histogramsHi} the goodness of the estimation of the idiosyncratic Hurst exponents. Thanks to synthetically generated data with $N=300$, we apply the calibration method to estimate the idiosyncratic Hurst exponents and compare them with the theoretical values. Here we simulate using the same Hurst exponents for all the idiosyncratic modes. One can see that the estimated Hurst exponents of the idiosyncratic modes lie around the chosen theoretical values: $H_i=0.03$ and $H_i=0.07$, which demonstrates a good calibration.

\section{Empirical Results}
\label{Section:empirical}

This section presents the results of the model calibration using empirical data.
We use historical daily open, high, low, and close price time-series of S\&P500 stocks from YahooFinance and implement the calibration method detailed in Section~\ref{sec:calibrationmethod}. \\

The estimation of the roughness of each log-volatility component the N-SfFM seeks to corroborate that the roughness inherent to the log-volatility of individual stocks predominantly stems from the residual log-volatility of the idiosyncratic components of stock returns. This idiosyncratic roughness vanishes when averaging stock returns to reconstruct index trajectories, whose roughness is then driven by the log-volatility factor roughness. Hence, for the  N-SfFM to align with the empirical results of Wu \textit{et al.} (see Figure~\ref{fig:wu_empirical_results}), we expect to find log-volatility components with the following Hurst exponents: $H\approx H_I\gtrsim 0.1 $ and for all stocks $i$, $H_{i}\approx 0$.\\

The first part of this section focuses on estimating the log-volatility factor roughness, confirming it aligns with that of the S\&P500 index, while the subsequent part addresses the estimation of the roughness of the idiosyncratic components.

\subsection{Estimating the factor roughness using daily market data}

\begin{figure}[H]
    \centering
    \includegraphics[width=0.8\linewidth]{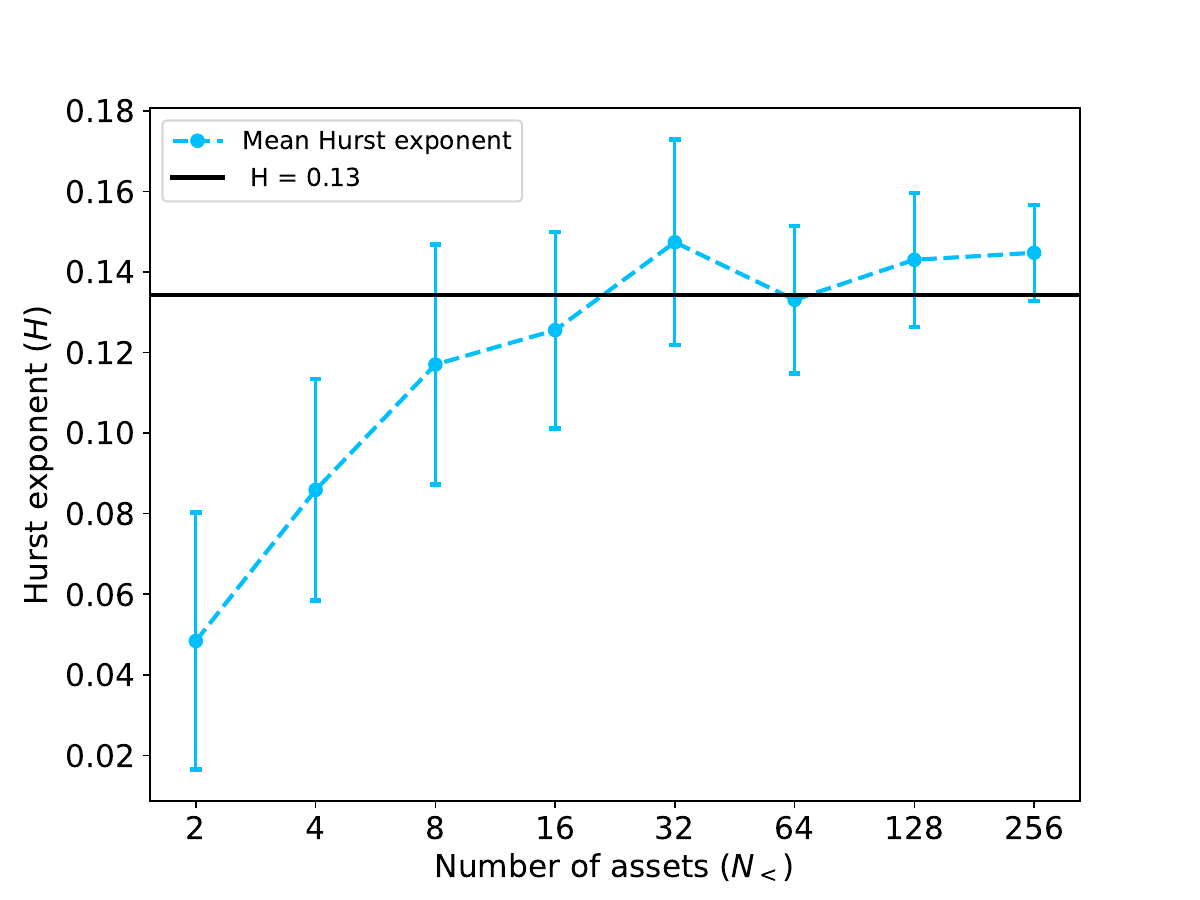} 
        \label{fig:subfig1}
    \caption{We plot using the S\&P500 market data for each asset number $N_s$ the mean value and the $95\%$ confidence interval of the Hurst exponent of the factor across multiple GMM time lag lengths via \textbf{blue} error bars, which we compare to the measured Hurst exponent of the S\&P500 index $H\simeq 0.13$,  represented by the horizontal \textbf{black} line.}
    \label{fig:fighurstfactor_emp}
\end{figure}

As noted in Section \ref{sec:calibrationmethod}, calculating the factor volatility directly from the daily volatility of the assets, as outlined in Eq.~\eqref{eq:qvfactor}, allows for the use of a volatility estimator with desirable characteristics, such as those related to log-volatility increment distribution and scale invariance properties \cite{mouti2023rough,garmanestimation}. An example of such an estimator is the Garman-Klass (GK) daily volatility estimator:
\begin{equation}\label{eq:GK-est}
    \hat{\sigma}_{GK}^2 = \frac{1}{2} \log^2\left(\frac{\text{High}}{\text{Low}}\right) - \left(2\log(2)-1\right) \log^2\left(\frac{\text{Close}}{\text{Open}}\right), \nonumber
\end{equation}
where High, Low, Open, Close stand respectively for the high, low, open, and close quotes over one day.\\\\
Using daily market data spanning from January 1, 2016, to January 1, 2020, we randomly selected 20 combinations of $N_s$ stocks from the S\&P500 index, with $N_s$ taking values from the set $\left(2^{k}\right)_{k\in \llbracket 1,8 \rrbracket}$. For each of these 20 combinations with a given $N_s$ number of stocks, we estimate the daily volatility of the factor, derived from the GK estimator of the stocks volatility through Eq.~\eqref{eq:qvfactor}.  Each of these 20 estimated factor volatilities, for a given $N_s$, then serves as an input in the Generalized Method of Moments (GMM), as introduced in \cite{wu2022rough}, to estimate the Hurst exponent of the factor, $H$. 
We compute the mean value of the Hurst exponent for the factor through 20 trials, with each trial performed with varying GMM time lags. These time lags are structured as $\left( \left\lfloor 2^{\frac{k}{4}} \right\rfloor\right)_{k\in \llbracket 0,Q \rrbracket}$ where the length $Q$ is derived from a uniform distribution $\mathcal{U}\left([28,40] \right)$. As a result, we obtain $20$ mean value estimations of the factor Hurst exponent $H$ (the mean value is computed from the 20-sized sample of the estimated $H$ that corresponds to each value of $Q)$  for each value of $N_s$. These estimations are compared with the Hurst exponent of the S\&P500 index, estimated using the GMM on the S\&P500 Garman-Klass estimator. Results are summarized in Figure~\ref{fig:fighurstfactor_emp}.

First, it should be noted that we retrieve the order of magnitude of the S\&P500 roughness ($H_I\gtrsim 0.1$) observed in previous works \cite{wu2022rough,gatheral2018volatility}. More importantly, Figure~\ref{fig:fighurstfactor_emp} supports the numerical experiments that demonstrate that the roughness of the factor reflects that of the S\&P500 index as $N_s$ increases. This further corroborates the idea that the roughness of indices stems from a common volatility component ($\Omega_t$) shared among individual stock dynamics. It also suggests that the ``very rough'' nature of individual stock volatility ($H_i\approx 0$) is likely attributable to their idiosyncratic component. An empirical investigation into the roughness of these components is provided in the next section.

Note also that Figure~\ref{fig:fighurstfactor_emp} reveals  that the effective Hurst exponent varies continuously as $N_s$ grows, from the multifractal behaviour characteristic of single stocks ($H_i \approx 0$)  to a rough behavior similar to that of indices ($H_I \approx 0.13$).

\subsection{Estimation of the roughness of idiosyncratic components}
\label{sec:EstimationRoughnessIdiosyncratic}
We now undertake the calibration of the N-SfFM, as introduced in Section~\ref{sec:calibrationmethod} and validated in Section~\ref{Section:numerics}, to estimate the roughness of the idiosyncratic components of stock volatility, denoted $({\omega}^{i})_{i\in \llbracket 1,243\rrbracket}$. 
The calibration is applied to empirical daily data from YahooFinance spanning from 01/01/1994 to 01/01/2026 for 243 US stocks that are constituents of the S\&P500 throughout the entire period.\\

To mitigate uncertainties and noise inherent to both the calibration process and the data, one can exploit the fact that the roughness of S\&P500 volatility has already been established in Ref.\cite{gatheral2018volatility}. This amounts to identifying the estimated daily factor volatility, $\Omega_t$, to the one of the S\&P500 index, approximated by the Garman-Klass estimator. It should be noted that the $\Omega$ time series derived after Step 3 of our calibration is 80\% correlated with the Garman-Klass estimator of the S\&P500 index. \\

\begin{figure}[H]
    \centering
    \includegraphics[width=\linewidth]{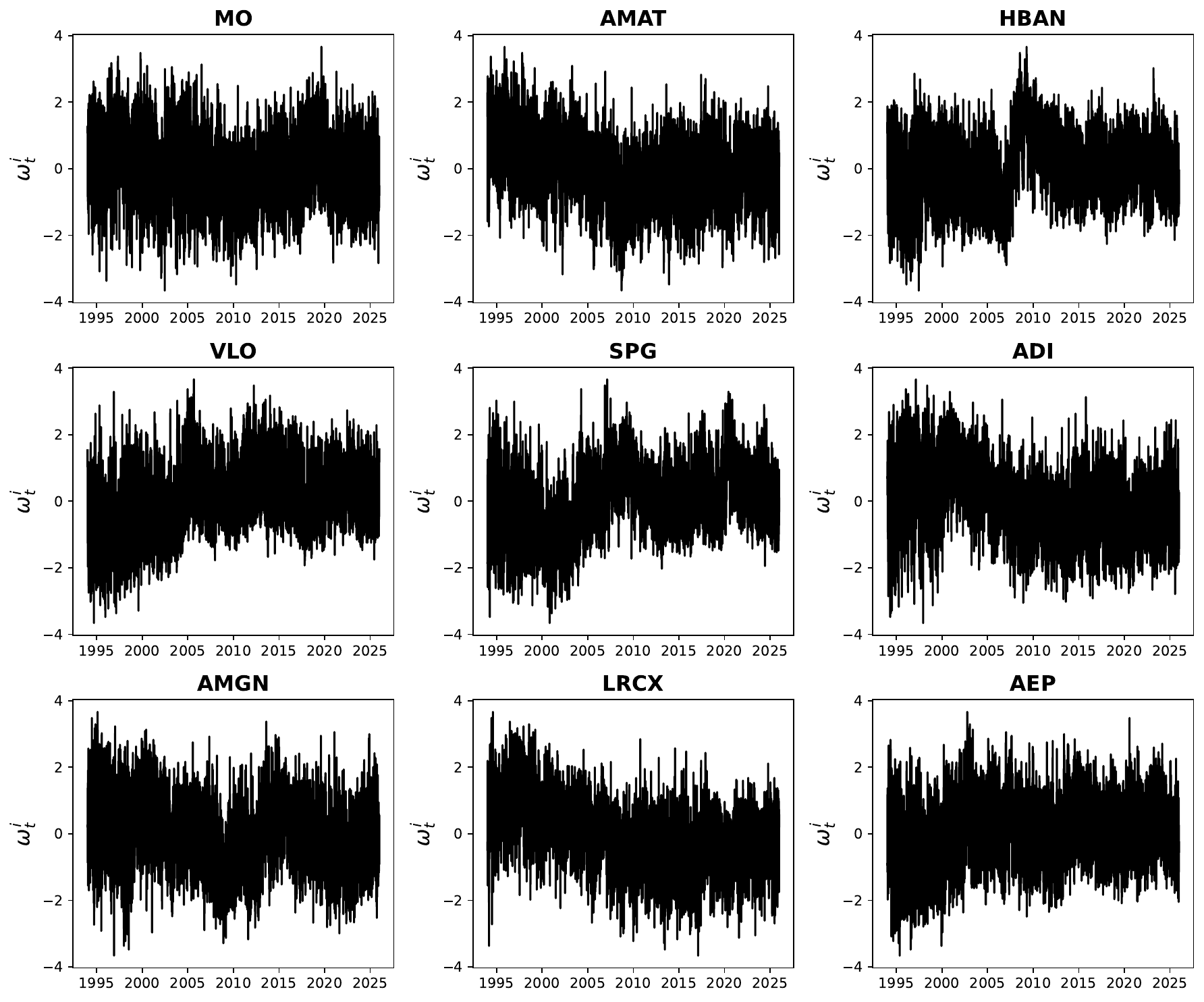}
    \caption{Idiosyncratic time series as estimated from market data of randomly selected S\&P500 single stocks. The complete nomenclature of the asset tokens is specified in Appendix \ref{app:assetnomenclature}.}
    \label{fig:omegaifrommktdata}
\end{figure}

Consequently, we employ two calibration approaches: the first involves estimating the factor log-volatility using the complete calibration method, while the second bypasses the second step of the calibration method by directly using the Garman-Klass estimate of the S\&P 500 log-volatility as the factor log-volatility. As an illustration, several time series of the idiosyncratic volatility component obtained through our calibration, denoted ${\omega}^{i}_t$, are displayed in Figure~\ref{fig:omegaifrommktdata}.



\begin{figure}
     \centering
     \begin{subfigure}[b]{0.45\textwidth}
         \centering
         \includegraphics[width=\textwidth]{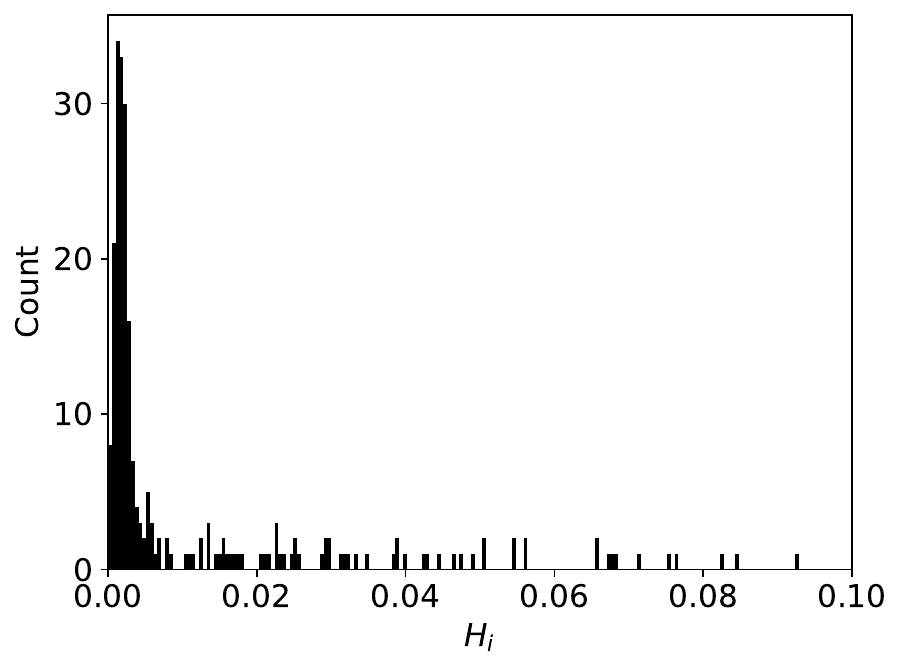}
         \caption{Deriving $\Omega_t$ from steps 1 and 2}
         \label{subfig:Hifrommktdata}
     \end{subfigure}
     \hfill
     \begin{subfigure}[b]{0.45\textwidth}
         \centering
         \includegraphics[width=\textwidth]{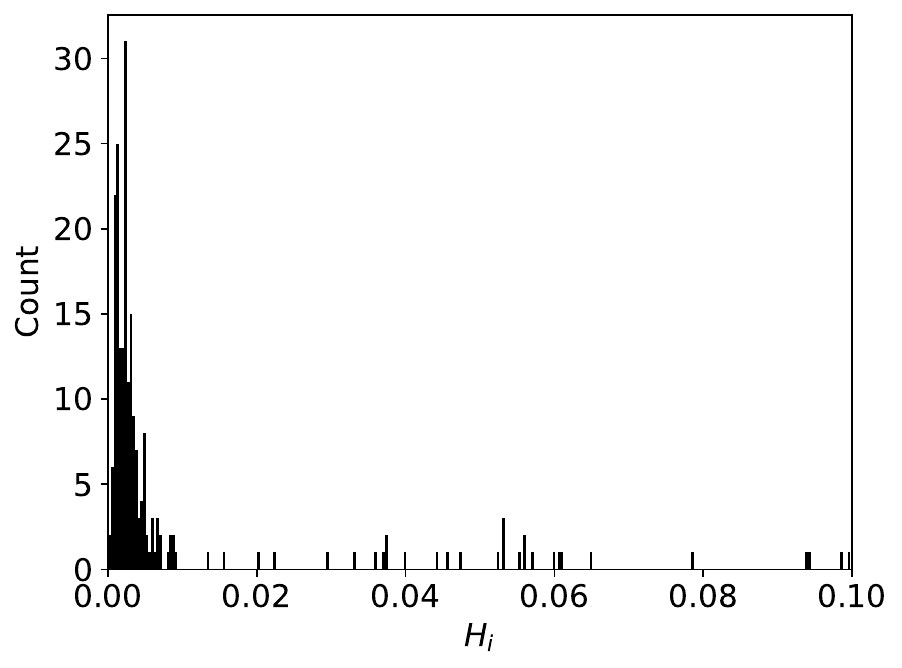}
         \caption{Sourcing $\Omega_t$ as the S\&P500 volatility}
         \label{subfig:Hifrommktdatasourcing}
     \end{subfigure}
     \caption{Estimated Hurst exponents $H_i$ from the estimated idiosyncratic time series of all the selected S\&P500 single stocks. Note that most of the $H_i$ are very close to zero.}
     \label{fig:calib_omega_i}
\end{figure}

Using the calibrated time series $\omega_t^{i}$ as inputs for the Generalized Method of Moments (GMM) of \cite{wu2022rough}, for all stocks, as outlined in the final step of the calibration Section~\ref{sec:calibrationmethod}, allows for the estimation of the idiosyncratic Hurst exponent for each stock. The resulting values are aggregated into empirical distributions, which are presented Figure~\ref{fig:calib_omega_i}. 

As expected, the results obtained from the full calibration method, depicted in Figure~\ref{subfig:Hifrommktdata}, and those obtained sourcing the S\&P500 volatility as the volatility factor, shown in Figure~\ref{subfig:Hifrommktdatasourcing}, are quite comparable. More importantly,
Figure~\ref{fig:calib_omega_i} illustrates that the Hurst exponents of the idiosyncratic volatility components, denoted $H_i$, are predominantly clustered around $H \simeq 0$. 
This demonstrates that the idiosyncratic volatility of individual stocks exhibits very high roughness compatible with a multifractal behavior ($H_i = 0$). It is noteworthy that such estimated values are not statistically distinguishable from those obtained for strictly vanishing Hurst exponent ($H = 0$). Numerical experiments using a genuine Multifractal Random Walk (MRW) path with $H = 0$ and a sample size comparable to our empirical S\&P 500 time series (approximately $3 \times 10^3$ data points; see Table 1 in \cite{wu2022rough}) reveal that mean value and the confidence interval are of the same order of magnitude as those observed in Figure~\ref{fig:calib_omega_i} where the means and standard deviations are of comparable magnitude (0.01 and 0.02 in Figure 8a; 0.02 and 0.03 in Figure \ref{fig:calib_omega_i}b).
Let us mention that a small subset of idiosyncratic residuals exhibits higher Hurst exponents ($H_i \gtrsim 0.1$). These outliers are omitted from Figure~\ref{fig:calib_omega_i}a and \ref{fig:calib_omega_i}b for visualization purposes. While this discrepancy may arise from empirical data noise, it could also indicate that certain stocks inherently possess smoother volatility profiles.

Combined with the results from the previous section, these findings substantiate our hypothesis: the roughness of individual stocks is driven by the multifractal idiosyncratic component. When single-stock trajectories are averaged to form the market index, this idiosyncratic contribution washes out, leaving only the less rough, common volatility component (the log-volatility factor) observable.

\section{Conclusion and Prospects}

Let us summarize our main results. This paper primarily aimed to demonstrate that the Nested Factor Model, incorporating Stationary fractional Brownian motions as log-volatilities, reproduces the somewhat surprising empirical findings of Wu \textit{et al.} \cite{wu2022rough}. Our framework was constructed on the intuition that the roughness of the combination of two signals with differing levels of roughness is dominated by the roughest signal. Hence, observing that Hurst exponents of single stocks are lower than those of indexes, we conjectured that their roughness originates from the idiosyncratic components of the stock dynamics, which average out in the dynamics of the indexes.   

The theoretical investigation of such a framework, leveraging the small intermittency approximation, allowed us to determine the regimes where such intuition holds, i.e. when the roughness of the idiosyncratic component is reflected in the individual trajectories of stocks. By simulating the proposed framework within these relevant regimes, we generated synthetic time series that match the empirical roughness observed in both stocks and indexes. Hence, we demonstrated that in addition to the enticing features of the Nested Factor Model established by Chicheportiche \textit{et al.} \cite{chicheportiche2015nested}, the N-SfFM also naturally accounts for different levels of roughness between stocks and indexes.

For greater simplicity and tractability, we limited the N-SfFM to its one-factor version, with no residual factor volatility. A natural extension of this work would be to investigate how our framework performs when incorporating additional factors, each with its own residual volatility modes, as originally introduced in \cite{chicheportiche2015nested}. These additional factors could represent different industry sectors, allowing one to investigate whether some industries exhibit particularly rough components. Accounting for the leverage and Zumbach effects is also an interesting direction to pursue.

\vfill
\textbf{Acknowledgment:} This work has received support from the French government, managed by the National Research Agency (ANR), under the "France 2030" program with reference "ANR-23-IACL-0008.

\newpage

\newpage
\section*{Appendices}
\begin{appendices}

\section{The M-factor Nested Factor model}
\label{app:MfactorNFM}

\begin{equation}\label{eq:MfactorNFM_definition}
    {\rm d}x^{i}_t = \sum_{k=0}^{M-1} \beta^{i}_k \, {\rm d}f^{k}_t + {\rm d}e^{i}_t \quad \text{with} \quad
    \begin{cases}
        {\rm d}f^{k}_t = \exp \left( \frac{F_k \Omega_t + r^{k}_t}{2} \right) {\rm d}W^{k}_t, \\
        {\rm d}e^{i}_t = \exp \left( \frac{A_i \Omega_t + \omega^{i}_t}{2} \right) {\rm d}B^{i}_t,
    \end{cases}
\end{equation}

where 
\begin{itemize}
\setlength\itemsep{0.01em}
    \item the $M$ time series $(f^{k}_{t})_{k\in\llbracket0,M-1\rrbracket,t>0}$ are the factors, serving as ``benchmarks" derived from the market. The series $f$ represents the market itself, and $({\rm d}f_{t})_t$ are frequently materialized as the returns of the S\&P500 index. Conversely, for $k>0$, ${\rm d}f^{k}_{t}$ may depict returns from a specific industrial sector. These time series $(f^{k}_{t})_{k\in\llbracket0,M-1\rrbracket,t>0}$ are uncorrelated;
    \item $\beta^{i}_k$ represents the exposure of asset $i$ to a factor ${\rm d}f^{k}$;
    \item the time series $({\rm d}e^{i}_{t})_i$ represent the residuals of each asset's returns, signifying the portion of the asset's returns not accounted for by market movements. These residuals are uncorrelated with the factors and are idiosyncratic to the asset itself;
    \item $\Omega_t$ is the log-volatility  stochastic \textit{factor} involved in the dynamics of both the factors $(f^{k}_{t})_{k\in\llbracket0,M-1\rrbracket,t>0}$ and the residuals $(e^{i}_{t})_i$. $r^{k}_t$ denotes the stochastic \textit{idiosyncratic} log-volatility of factor $f^{k}$, while $\omega^{i}_t$ is the stochastic \textit{idiosyncratic} log-volatility of residual $e^{i}$;
    \item parameters $\gamma_i$'s and $F_k$'s quantify the contribution of each factor volatility mode. As discussed in\cite{chicheportiche2015nested}, they can be estimated using correlation functions (denoted $C^{ff}$ and $C^{rr}$ in \cite{chicheportiche2015nested}, note that, their dominant eigenvector provide a reliable initial estimate);
    \item finally, $W^{k}_t$ and $B^{i}_t$ are Brownian motions, all independent with each other.
\end{itemize}    


\section{The Log S-fBM model}
\label{app:appendixLog S-fBM}
In this Appendix, we will use the same notations as in \cite{wu2022rough}.\\\\
The Log S-fBM model is a stochastic volatility model defined introduced in \cite{wu2022rough} whose dynamic is:
\begin{equation}
\label{eq:logsfBM}
        \begin{cases*}
      dX_t = e^{\frac{\omega_{H,T}(t)}{2}} dB_t\\*
      X_0=x_0 
    \end{cases*},
    \end{equation}

  where:
    \begin{itemize}
   
    \item $\omega_{H,T}(.)$ is the S-fBM process, a stationary Gaussian process with the following autocovariance function in Eq \ref{eq:covSfBM}, and mean:
 \begin{equation}
        \mu_{H,T}=-\frac{\lambda^2}{4H(1-2H)};
    \end{equation}
         \item $B$ is a standard Brownian motion independent of $\omega_{H,T}$;
        \item $\lambda>0$, $H\in ]0,1]$, and $T>0$ are respectively the intermittency parameter, the Hurst exponent and the decorrelation parameter.
    \end{itemize}
By construction, for any $\tau>0$, the S-fBM increment process $\left(\delta_{\tau}\omega_{H,T}(t):=\omega_{H,T}(t+\tau)-\omega_{H,T}(t)\right)_t$ is a stationary Gaussian process, centered and  with variance $v\left(\tau,H\right)=\E\left(|\delta_{\tau}\omega_{H,T}(t)|^2 \right)$ expressed as:
\begin{eqnarray}
\label{eq:varianceSfBMincrementvariance}
    v\left(\tau,\nu,H\right)=\nu^2\left(\frac{\tau}{T}\right)^{2H},
\end{eqnarray}
where $\nu^2=\frac{\lambda^2}{H(1-2H)}$, which is a simple way to parametrize the S-fBM variance:
\begin{eqnarray}
    \forall t\geq 0,\tab  \text{Var}\left(\omega_{H,T}(t) \right)=\frac{\nu^2}{2}. \nonumber
\end{eqnarray}
For any interval $K\subset \R_+$ of length $|K|$ such that $|K|\leq T$ and any sub-interval $ I\subset K$, 
let us introduce the integrated process:
\begin{eqnarray}                              
\label{eq:integratedlogvol}
    \Upsilon(I):=\frac{1}{\lambda}\int_I \left(\omega_{H,T}(u)-\mu_{H,T}\right) du,
\end{eqnarray}
and, for any $t >0$ and $0<\Delta\leq T$,
\begin{eqnarray}
   \Upsilon_{\Delta}(t):=\Upsilon([t,t+\Delta]).
\end{eqnarray}

In \cite{wu2022rough} it is established that:
\begin{prop}
\label{prop:propositioncovOmega}
   The process 
   $\Upsilon_{\Delta}$ is a stationary centered Gaussian process, with the following covariance structure:

    \begin{eqnarray}
   \label{eq:autocovOmegaDelta}
    \forall (t,s)\in\R_{+}^2,\tab  
    \text{Cov}\left(\Upsilon_{\Delta}(t),\Upsilon_{\Delta}(s) \right)=\frac{\Delta^2}{2H(1-2H)}-\Delta^2\left(\frac{\tau}{T}\right)^{2H}\tilde{g}_{H}\left(\frac{\Delta}{\tau}\right), \nonumber \\
    \end{eqnarray}
   where $\tau=|t-s|$ and 
    $\tilde{g}_{H}\left(z\right):=\frac{|1+z|^{2H+2}+|1-z|^{2H+2}-2}{2Hz^2(1-(2H)^2)(2H+2)}$.
\end{prop}
\textbf{\textit{Proof}}
We fix $\Delta>0$.\\
By definition, $ \left(\omega_{H,T}(t)-\mu_{H,T}\right)_t$ is a centered Gaussian process. Consequently, $\Omega_{H,T,\Delta}$ is a centered Gaussian process too.\\\\
We suppose that $t>s$ as they play symmetric roles.\\
    We have:
    \begin{eqnarray}
        \text{Cov}\left(\Upsilon_{\Delta}(t),\Upsilon_{\Delta}(s) \right)= \frac{1}{\lambda^2}\int_t^{t+\Delta}\int_s^{s+\Delta}\text{Cov}\left(\omega_{H,T,\Delta}(t),\omega_{H,T,\Delta}(s) \right)dudv \nonumber
    \end{eqnarray}
    By definition, we have :
    \begin{eqnarray}
        \frac{1}{\lambda^2}\int_t^{t+\Delta}\int_s^{s+\Delta}\text{Cov}\left(\omega_{H,T}(t),\omega_{H,T}(s) \right)dudv= \frac{1}{2H(1-2H)}\int_t^{t+\Delta}\int_s^{s+\Delta}\left(1-\left(\frac{|u-v|}{T}\right)^{2H}\right)dudv \nonumber
    \end{eqnarray}
    After some algebra, one has:
    \begin{eqnarray}
        \frac{1}{\lambda^2}\int_t^{t+\Delta}\int_s^{s+\Delta}\text{Cov}\left(\omega_{H,T}(t),\omega_{H,T}(s) \right)dudv=\frac{\Delta^2}{2H(1-2H)}-\frac{|\tau+\Delta|^{2H+2}- 
        \tau^{2H+2}}{2HT^{2H}(1-(2H)^2)(2H+2)}\nonumber \\
        -\frac{1}{2HT^{2H}(1-(2H)^2)}\left(\int_t^{s+\Delta}|u-s-\Delta|^{2H}(u-s-\Delta) du+\int_{s+\Delta}^{t+\Delta}|u-s-\Delta|^{2H}(u-s-\Delta) du\right) \nonumber
    \end{eqnarray}
    \\
    By integrating the remaining term, we obtain:
    \begin{eqnarray}
        \frac{1}{\lambda^2}\int_t^{t+\Delta}\int_s^{s+\Delta}\text{Cov}\left(\omega_{H,T}(t),\omega_{H,T}(s) \right)dudv=\frac{\Delta^2}{2H(1-2H)}-
        \frac{|\tau+\Delta|^{2H+2}-\tau^{2H+2}+|\tau-\Delta|^{2H+2}-\tau^{2H+2}}{2HT^{2H}(1-(2H)^2)(2H+2)}\nonumber
    \end{eqnarray}
    Which leads to:
    \begin{eqnarray}
        \frac{1}{\lambda^2}\int_t^{t+\Delta}\int_s^{s+\Delta}\text{Cov}\left(\omega_{H,T}(t),\omega_{H,T}(s) \right)dudv=\frac{\Delta^2}{2H(1-2H)}-
        \frac{|\tau+\Delta|^{2H+2}-2\tau^{2H+2}+|\tau-\Delta|^{2H+2}}{2HT^{2H}(1-(2H)^2)(2H+2)}\nonumber
    \end{eqnarray}
   
Thus, $\Upsilon_{\Delta}(.)$ is a stationary Gaussian process with the claimed autocovariance structure. \\
    \tab\tab\tab\tab\tab\tab\tab\tab\tab\tab\tab\tab\tab\tab\tab\tab\tab\tab\tab\tab\tab\tab\tab\tab\tab\tab\tab\tab\tab\tab\tab\tab\tab\tab $\blacksquare$\\
\begin{remark}
    We can derive the autocovariance $C_{\Upsilon}\left(\Delta,\tau\right):=\text{Cov}\left(\frac{\Upsilon([0,\Delta])}{\Delta},\frac{\Upsilon([\tau,\tau+\Delta])}{\Delta}\right)$ for any $\Delta>0$ as follows:
\begin{eqnarray}
\label{eq:autocovCdeltatautheo}
    C_{\Upsilon}\left(\Delta,\tau\right)=\frac{1}{2H(1-2H)}-\left(\frac{\tau}{T}\right)^{2H}\tilde{g}_{H}\left(\frac{\Delta}{\tau}\right)
\end{eqnarray}
\end{remark}

\section{The small intermittency approximation and intermittency bias estimation}
\label{app:smallintermittencyapproximationconsistency}
In this section, we first assess the accuracy of the small intermittency approximation used to compute the covariance of the S-fBM. We then demonstrate how this approximation of the moments allows one to control the bias in GMM estimation.

\subsection{Accuracy of the small intermittency approximation}
\label{app:accuracysmallintermittencyapproximationconsistency}
In the sequel, let $K\subset \R_+$ be an interval such that $|K|\leq T$.\\\\
We consider the Multifractal random measure associated to the Log S-fBM model (\ref{eq:logsfBM}) as defined in Eq .~\eqref{eq:NFM_MRM_M_def}:
\begin{eqnarray}
\label{eq:mrm}
    \forall I\subset K,\tab M_{H,T}(I):=\int_Ie^{\omega_{H,T}(u)}du.
\end{eqnarray}
It can be seen as its integrated volatility over $I$. As well as its integrated counterpart over a period of length $\Delta$:
\begin{eqnarray}
\label{eq:mrmdelta}
\forall t \geq 0,\tab
     M_{\Delta}(t):=M_{H,T}([t,t+\Delta]).
\end{eqnarray}

This appendix showcases the consistency of considering the small intermittency approximation up to the order one in $\lambda^2$ as used in the Log S-fBM calibration method of Wu \textit{et al.} in Section 4 in \cite{wu2022rough}.\\
Our goal is to control the error, in the small intermittency limit of the first order approximation for any arbitrary $I,J$ in $K$:
\begin{equation}
     \mathrm{Cov} \left[ \ln \left(\frac{M_{H,T}(I)}{|I|}\right) \! \! , \ln \left( \frac{M_{H,T}(J)}{|J|}\right) \right] \simeq  \lambda^2 \text{Cov}\left( \Upsilon(I),\Upsilon(J) \right) =
     \frac{\lambda^2}{|I||J|}\int_{I\times J} \! \! \! \text{Cov}\left(\frac{1}{\lambda}\omega_{H,T}(t),\frac{1}{\lambda}\omega_{H,T}(s))\right){\rm d}t{\rm d}s
\end{equation} 
where $\Upsilon(I)$ is defined in \eqref{eq:integratedlogvol} in previous Appendix.
We consider the case $I =[0,\Delta]$ and $J = [\tau,\tau+ \Delta]$. One has:
\begin{eqnarray*}
 \frac{1}{\lambda} \ln \Big( \frac{M_{\Delta}(t)}{\Delta} \Big) & = & \frac{1}{\lambda} \ln \Big( e^{\lambda \frac{\Upsilon_\Delta(t)}{\Delta}} \frac{1}{\Delta}\int_t^{t+\Delta} e^{\omega_{H,T}(u) - \lambda \frac{\Upsilon_\Delta(t)}{\Delta}} du  \Big) \\ 
  & = & \frac{\Upsilon_\Delta(t)}{\Delta}  +  \frac{1}{\lambda} \ln \Big( \frac{1}{\Delta}\int_t^{t+\Delta} e^{\omega_{H,T}(u) - \lambda \frac{\Upsilon_\Delta(t)}{\Delta}} du \Big)
 \end{eqnarray*}
We perform the second order series expansion of the latter quantity. We refer the reader to Kozhemyak \textit{et al.} \cite{bacry2008log}  for the technical details and assumptions .
Using the same notation as in Notation 2 in \cite{bacry2008log}, we have to the first order in $\lambda$:
\begin{eqnarray*}
 \frac{1}{\lambda} \ln \Big( \frac{M_{\Delta}(t)}{\Delta} \Big) & \stackrel{\lambda^2}{\simeq} &  \frac{\Upsilon_\Delta(t)}{\Delta}  +  \frac{\lambda}{2 \Delta} \int_t^{t+\Delta} \left(z_u - \frac{\Upsilon_\Delta(t)}{\Delta}\right)^2 du  \\
 & \stackrel{\lambda^2}{\simeq} & \frac{\Upsilon_\Delta(t)}{\Delta} + \frac{\lambda}{2 \Delta}  \left(
  \int_t^{t+\Delta} z_u^2 \; du - \frac{\Upsilon_\Delta(t)^2}{\Delta} \right)
\end{eqnarray*}
where we set $z_u = \frac{\omega_{H,T}(u)-\mu_{H,T}}{\lambda}$.\\
Let 
$$
  \Theta_\Delta(t) =  \int_t^{t+\Delta} z_u^2 \; du - \frac{\Upsilon_\Delta(t)^2}{\Delta} 
$$
and the associated autocovariance function: $$
C(\tau) = \text{Cov}\left( \Theta_\Delta(t),\Theta_\Delta(t+\tau)\right).
$$
If one uses previous expression, one has using Wick's theorem (see \cite{wick1950collision}),
\begin{eqnarray}
    \forall \tau\geq 0,\tab \text{Cov} \Big(\frac{\Upsilon_\Delta(t)}{\Delta}, \Theta_\Delta(t+\tau) \Big) = 0\nonumber
\end{eqnarray}

and it results that:
\begin{eqnarray}
    \text{Cov}\left( \frac{1}{\lambda} \ln \Big( \frac{M_{\Delta}(t)}{\Delta} \Big),
\frac{1}{\lambda} \ln \Big( \frac{M_{\Delta}(t+\tau)}{\Delta} \Big) \right) = 
\text{Cov}\left( \frac{\Upsilon_\Delta(t)}{\Delta}, \frac{\Upsilon_\Delta(t+\tau)}{\Delta} \right)
+ \frac{\lambda^2}{4\Delta^2} C(\tau) + o\left(\lambda^2 \right).
\end{eqnarray}

\begin{figure}[H]
  \center
    \includegraphics[width=0.8\linewidth] {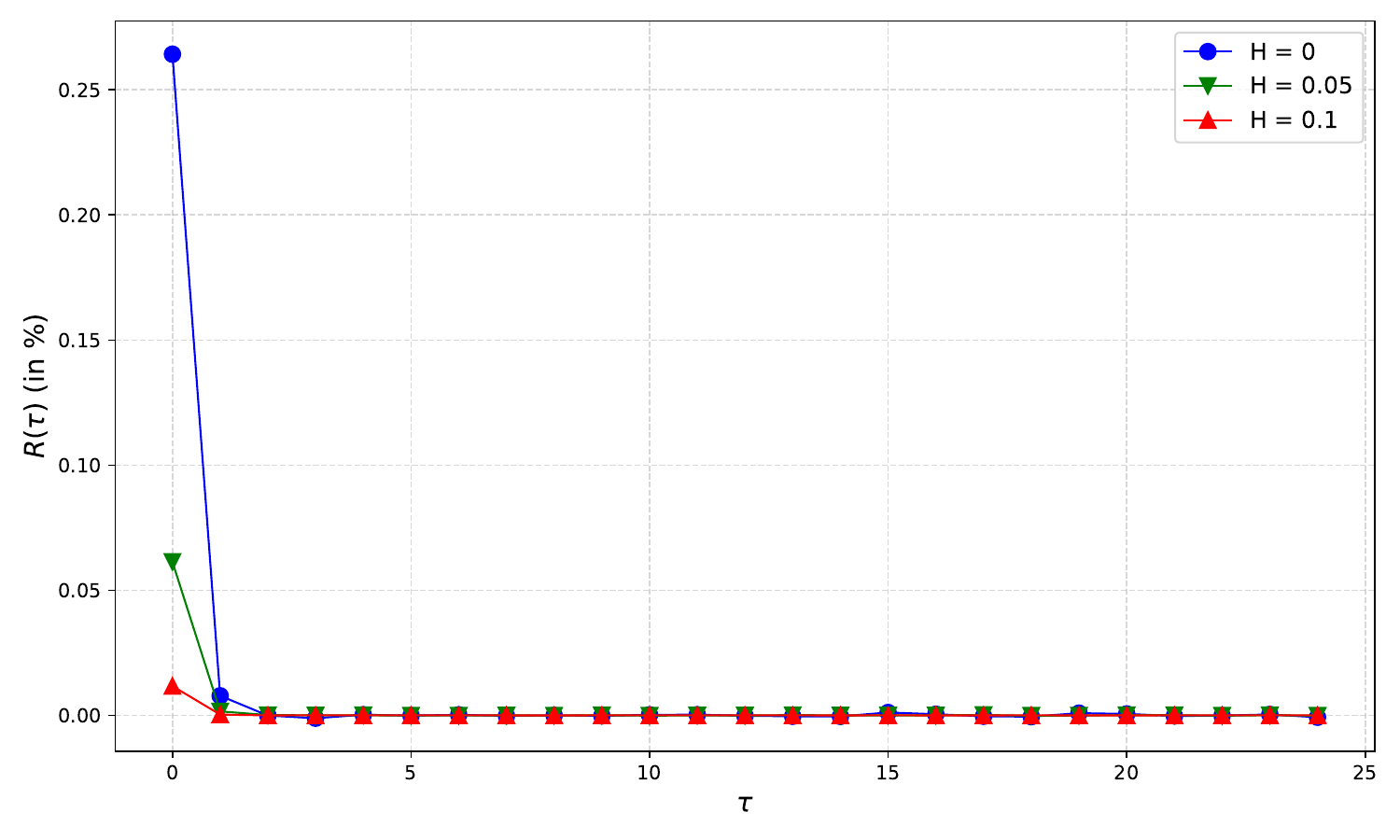}\hfill
    
    \caption{Numerical estimation of $R(\tau)$ for $H=0,0.05$ and $0.1$ for $\lambda^2=0.05$, $T=5. \; 10^3$ and $\Delta =1$. One can see that as soon as $\tau \geq 2$, $R(\tau)$ is negligible.} 
    \label{fig:RSI}
\end{figure}

In order to assess the small intermittency approximation quality one may estimate the following ratio: 

$$ R(\tau) = \frac{\lambda^2 C(\tau)}{4 \;\text{Cov}\left(\Upsilon_{\Delta}(0),\Upsilon_{\Delta}(\tau) \right)} \; 
$$
For the sake of simplicity, we consider the case $H = 0$ and $\tau = 0$, where the analytical computation can be carried out explicitly. In that case, one has
$$
 C(0) = \mathrm{Var}\left(\int_0^\Delta z^2_u \; du \right) +   \mathrm{Var}\left( \frac{\Upsilon_\Delta(0)^2}{\Delta}  \right) - 2 \; \mathrm{Cov}\left( \int_0^\Delta z^2_u \; du , \frac{\Upsilon_\Delta(0)^2}{\Delta} \right)
$$
Thanks to Wicks theorem, it can be shown that previous expression reduces to 
$$
 C(0) = V_1 + 2 V_2^2 - 2 V_3
$$
with
\begin{eqnarray*}
 V_1 & = & 4 \int_0^\Delta  \left(\Delta-s \right) \ln^2 \left( \frac{T}{s} \right) \; ds  = \Delta^2 \left( 2 \ln^2 \left( \frac{T}{\Delta} \right) + 6 \ln \left( \frac{T}{\Delta} \right) + 7 \right)
 \\
 V_2 & = & \frac{2}{\Delta} \int_0^\Delta (\Delta - s)  \ln \left( \frac{T}{s} \right) \; ds  = \Delta \left( \ln \left( \frac{T}{\Delta} \right) + \frac{3}{2} \right) \\
 V_3 & = & \frac{2}{\Delta} \int_0^\Delta \int_0^\Delta \int_0^\Delta \ln \left( \frac{T}{|s-v|} \right) \ln \left( \frac{T}{|s-u|} \right) ds \;  du \;  dv 
 =  \Delta^2 \left( 2 \ln^2 \left( \frac{T}{\Delta} \right) + 6 \ln \left( \frac{T}{\Delta} \right) + \frac{17}{3}-\frac{\pi^2}{9} \right) \; .
\end{eqnarray*}
This leads to the final expression for $C(0)$:
$$
  C(0) = \left( \frac{1}{6} + \frac{2 \pi^2}{9} \right) \Delta^2 \approx 2.36 \; \Delta^2
$$
Since
$$
  \mathrm{Var} \left(\Upsilon_{\Delta}\right) = V_2 \; , 
$$
for $T = 5000$ and $\Delta =1$, $V_2 \approx 10$, and we finally estimate, 
when $\lambda^2 = 0.05$,
$$ 
 R(0) \approx 0.003 = 0.3 \% 
$$
which is rather small. The exact computation of $R(\tau)$ for other values of $H$ and for any lag $\tau$ is trickier. As illustrated in Figure \ref{fig:RSI}, for any $H$, $C(\tau) \approx 0$ when $\tau \geq \Delta$. Moreover, one can see that
that $R(\tau)$ decreases as $H$ increases, as estimated previously, is the worst case. Therefore, the error by approximating  
$\text{Cov}\left( \frac{1}{\lambda} \ln \Big( \frac{M_{\Delta}(t)}{\Delta} \Big),
\frac{1}{\lambda} \ln \Big( \frac{M_{\Delta}(t+\tau)}{\Delta} \Big) \right)$ by
$\text{Cov}\left( \frac{\Upsilon_\Delta(t)}{\Delta}, \frac{\Upsilon_\Delta(t+\tau)}{\Delta} \right)$ appears to be smaller than 0.3 \%.

In Figure \ref{fig:RSI}, we followed for the case $H=0$ the simulation method explained in \cite{bacry2008log} with an arbitrary small cut-off scale $\ell$ (see Eq.~\eqref{eq:logcorrelprocesswithcutoff}). For strictly positive Hurst exponents, the simulation method is similar to what has been performed in Section \ref{Section:numerics} as well as the numerical experiments of \cite{wu2022rough}.

\subsection{The bias of the estimated intermittency coefficient}
\label{app:intermittencybias}
The GMM methodology employed to estimate the log-S-fBm parameters follows the framework originally proposed in \cite{wu2022rough}. For a sequence of time lags $\Tau \subset \mathbb{N}^q$, the estimator is constructed from the empirical moments:
\[
K_N(\theta) = \left(\Hat{C}_{\text{log}}(k\Delta) - \lambda^2 C_{\Upsilon}(\Delta, k\Delta)\right)_{k \in \Tau},
\]
where $\Hat{C}_{\text{log}}$ is the empirical covariance function of the log-increments and $C_{\Upsilon}$ is the analytical counterpart derived under the small intermittency approximation as discussed previously. 

In this setting, the standard correctly specified moment condition $E[K_N(\theta_0)] = 0$ is replaced by a population moment $K(\theta) = \mathbb{E}[K_N(\theta)]$ that satisfies:
\begin{equation}\label{eq:misspec_bound}
K(\theta_0) = g(\theta_0), \quad \text{where} \quad \|g(\theta_0)\| = o(\lambda_0^2).
\end{equation}
This formulation aligns with the ``non-local'' misspecification framework discussed in \cite{hall2003large, hall2005generalized}, where the expected empirical moments depend systematically on the parameters. This is distinct from the ``local'' misspecification analyzed by \cite{armstrong2021sensitivity}, which assumes the moment conditions vanish as $N \to \infty$.

The GMM estimator $\hat{\theta}_N$ is defined as the minimizer of the sample objective function:
\[
Q_N(\theta) = K_N(\theta)' W_N K_N(\theta),
\]
where $\{W_N\}$ is a sequence of positive definite weighting matrices such that $W_N \xrightarrow{p} W$ for some positive definite matrix $W$. Under standard regularity conditions \citep{hansen1982large, hall2005generalized}, $\hat{\theta}_N$ converges in probability to the pseudo-true parameter $\theta^*$:
\[
\theta^* = \arg\min_{\theta \in \Theta} Q(\theta), \quad \text{where} \quad Q(\theta) = K(\theta)' W K(\theta).
\]

To control the asymptotic bias, we analyze the population first-order condition $D(\theta^*)' W K(\theta^*) = 0$, where $D(\theta) = \frac{\partial K(\theta)}{\partial \theta'}$ denotes Jacobian matrix. Assuming $D(\theta_0)$ has full column rank, we perform a first-order Taylor expansion of $K(\theta^*)$ around the true parameter $\theta_0$:
\[
K(\theta^*) = K(\theta_0) + D(\bar{\theta})(\theta^* - \theta_0),
\]
where $\bar{\theta}$ lies on the line segment connecting $\theta^*$ and $\theta_0$. Substituting $K(\theta_0) = g(\theta_0)$ and the expansion into the first-order condition, we obtain:
\begin{equation}\label{eq:firstorderbias_restored}
\theta^* - \theta_0 = -(D(\theta^*)' W D(\bar{\theta}))^{-1} D(\theta^*)' W g(\theta_0).
\end{equation}
Notice that the same kind of computation was undertaken in the context of sensitivity analysis of the GMM estimates in \cite{andrews2017measuring}. By the continuity of $D(\theta)$ and the consistency of the approximation, \eqref{eq:firstorderbias_restored} implies $\|\theta^* - \theta_0\| = O(\|g(\theta_0)\|)$. Leveraging the approximation bound from \eqref{eq:misspec_bound}, we conclude:
\begin{equation}
\hat{\lambda}^2_N \xrightarrow[N\to\infty]{\Prob} \lambda_0^2 + o(\lambda_0^2).
\end{equation}
This confirms that the estimator remains asymptotically valid within the small-intermittency regime, as the approximation error is dominated by the magnitude of the intermittency parameter itself.

\subsection{Discussion on regularity conditions and full rankedness of $D$  }
\label{app:fullrankD}
The development of Section \ref{app:intermittencybias} rely mainly on regularity condition of the objective function $K$.
The assumptions in \citep{hansen1982large} are standard well-behavedness conditions for GMM: a bounded parameter space with the true value safely interior, a moment function continuously differentiable with bounded derivatives. These hold naturally here. Choosing $\Theta\subset ]0,\tfrac12 [\times\mathbb R^*$ compact and bounded away from the singular values $H\in \{0,\tfrac12\}$ keeps $\theta^*$ interior. $C_\Upsilon(\Delta,\tau)$ is built from smooth closed-form expressions.\\

In the sequel, we justify the full rankedness assumption of $D$ being the Jacobian matrix of $K$ as this is also behind the regularity conditions of \citep{hansen1982large, hall2005generalized}.\\

In the sequel, we denote: \[
C_\Upsilon(\Delta,\tau) = A(H) - x(\tau)\,\tilde g_H(z),
\qquad
A(H) := \frac{1}{2H(1-2H)}, \quad
x(\tau) := \left(\frac{\tau}{T}\right)^{2H}, \quad
z := \frac{\Delta}{\tau},\quad L(\tau) := \ln(\tau/T)
\]
One has, 
\[
\partial_\lambda K(\theta) = -2\lambda\, C_\Upsilon(\Delta,k\Delta).
\]
Which means that:
\[
\mathbf{c}_\lambda := \big(\partial_\lambda K(\theta)\big)_{k\in\mathcal T} = -2\lambda\,\big(C_\Upsilon(\Delta,k\Delta)\big)_{k \in \mathcal T}.
\]
On the other hand, applying the chain rule gives:
\[
\partial_H\big(x(\tau)\tilde g_H(z)\big) = 2L(\tau)\,x(\tau)\tilde g_H(z) + x(\tau)\,\partial_H \tilde g_H(z).
\]
Since $C_\Upsilon(\Delta,\tau) = A(H) - x(\tau)\tilde g_H(z)$, we have $x(\tau)\tilde g_H(z) = A(H) - C_\Upsilon(\Delta,\tau)$, and substituting:
\[
\partial_H C_\Upsilon(\Delta,\tau)
= A'(H) - 2L(\tau)\big[A(H)-C_\Upsilon(\Delta,\tau)\big] - x(\tau)\partial_H\tilde g_H(z).
\]
Rearranging,
\begin{eqnarray}
\partial_H C_\Upsilon(\Delta,\tau) = 2L(\tau)\, C_\Upsilon(\Delta,\tau) + R(H,\tau),\nonumber
\end{eqnarray}
where the remainder is
\begin{eqnarray}
R(H,\tau) := A'(H) - 2L(\tau)A(H) - x(\tau)\,\partial_H\tilde g_H(z).
\nonumber
\end{eqnarray}
Therefore
\[
\partial_H K(\theta) = -\lambda^2\Big[2L(k\Delta)\,C_\Upsilon(\Delta,k\Delta) + R(H,k\Delta)\Big],
\]
and the second Jacobian column is
\[
\mathbf{c}_H := \big(\partial_H K(\theta)\big)_{k\in\mathcal T}
= -\lambda^2\Big(2L(k\Delta)\,C_\Upsilon(\Delta,k\Delta) + R(H,k\Delta)\Big)_{k\in\mathcal T}.
\]

Since $\mathbf c_\lambda \neq \mathbf 0$, claiming that $J$ is not full rank forces $\mathbf{c}_H = c\,\mathbf{c}_\lambda$ for some scalar $c \in \mathbb{R}$, i.e.
\[
-\lambda^2\Big[2L(k\Delta)C_\Upsilon(\Delta,k\Delta) + R(H,k\Delta)\Big] = -2\lambda c\, C_\Upsilon(\Delta,k\Delta), \qquad \forall\, k \in \mathcal T.
\]
Setting $c' := 2c/\lambda$, this simplifies to
\begin{eqnarray}
\forall\, k \in \mathcal T, \qquad \big(2L(k\Delta) - c'\big)\,C_\Upsilon(\Delta,k\Delta) = -R(H,k\Delta).\nonumber
\end{eqnarray}
This is equivalent to
\begin{eqnarray}
\forall\, k \in \mathcal T, \qquad  \Psi(k\Delta) := 2L(k\Delta) - \frac{R(H,k\Delta)}{C_\Upsilon(\Delta,k\Delta)} = c'.
\end{eqnarray}
By direct computations applying the quotient rule for a given $\tau>0$, 
\begin{eqnarray}
    \Psi'(\tau) = \frac{2}{\tau} - \frac{R'(\tau)C_\Upsilon(\tau) - R(\tau)C_\Upsilon'(\tau)}{C_\Upsilon(\tau)^2}
= \frac{1}{\tau}\left[2 - \frac{\widetilde R(\tau)C_\Upsilon(\tau) - R(\tau)\widetilde C(\tau)}{C_\Upsilon(\tau)^2}\right].
\end{eqnarray}
where $C_\Upsilon'(\tau) = \tfrac{1}{\tau}\widetilde C(\tau)$, $R'(\tau)=\tfrac1\tau \widetilde R(\tau)$ and
\[
\widetilde C(\tau) := x(\tau)\big[zg'(z)-2Hg(z)\big], \qquad
\widetilde R(\tau) := -2A(H) - 2Hx(\tau)\partial_H g(z) + zx(\tau)\partial_z\partial_H g(z).
\]
Thus $\Psi'(\tau)=0$ at a given $\tau>0$ if and only if the algebraic identity
\begin{eqnarray}
    2\,C_\Upsilon(\Delta,\tau)^2 = \widetilde R(\tau)\,C_\Upsilon(\Delta,\tau) - R(H,\tau)\,\widetilde C(\tau)
\end{eqnarray}

holds. Since the left side is quadratic in $C_\Upsilon$, the latter identity has no structural reason to hold for generic $(H,\Delta,T,\tau)$.

\section{Proof of Theorem \ref{theo:variancesumsmallintermit}}
\label{app:proofoftheovariancesumsmallintermit}
This Section is devoted to the proof of Theorem \ref{theo:variancesumsmallintermit} using the notations of Section \ref{sec:SIA}.\\\\
We consider here any sub-intervals $(I,J)\subset K^2$. We introduce the following notations:

\begin{eqnarray}
    \begin{cases}
         C^{log}_{\Sigma}(I,J):=\E\left(\ln\left(\frac{\Sigma\left(I\right)}{|I|}\right)\ln\left(\frac{\Sigma\left(J\right)}{|J|}\right)\right)\\
         C^{log}(I,J):=\E\left(\ln\left(\frac{M\left(I\right)}{|I|}\right)\ln\left(\frac{M\left(J\right)}{|J|}\right)\right)\\
          C_i(I,J):=\E\left(\frac{\Upsilon_{i}(I)}{|I|}\frac{\Upsilon_{i}(J)}{|J|}\right), i\in \llbracket 1,d \rrbracket
    \end{cases}
    \nonumber
    \end{eqnarray}
where $\Upsilon_{i}(I):=\frac{1}{\lambda_i}\int_I \left(\omega^{i}_u-\mu_i\right) du$.\\

Beforehand, we need this intermediary result.
\begin{lem}
\label{thm:smallinterautocovinterval}
For any sub intervals $I$ and $J$ of $K$, the following approximations holds:

\begin{enumerate}[label=\alph*)]  
    \item 
\begin{eqnarray}
      C^{log}(I,J)-C^{log}(I,I)=\sum_{i=1}^d a_i^2\lambda_{i}^2\left(C_i(I,J)-C_i(I,I)\right) + o\left(\left|\left|\LL\right|\right|^2\right),
\end{eqnarray}
  \item 
  \begin{eqnarray}
      C^{log}_{\Sigma}(I,J)-C^{log}_{\Sigma}(I,I)=\sum_{i=1}^d b_i^4 \lambda_{i}^2\left(C_i(I,J)-C_i(I,I)\right) + o\left(\left|\left|\LL\right|\right|^2\right),
\end{eqnarray}
\end{enumerate}

where $||.||$ is the $\mathcal{L}^2$ norm in $\R^d$ and the tuples $\left(a_i\right)_{i\in \llbracket 1,d \rrbracket}$ and $\left(b_i\right)_{i\in \llbracket 1,d \rrbracket} $ are defined in the Section \ref{sec:SIA}.
\end{lem}

\textbf{\textit{Proof}}
\begin{enumerate}[label=\alph*)]  
\item 
By straightforward computations, we have for any subintervals $I$ and $J$ of $K$ of length $T$:
\begin{eqnarray}
     \E\left(\left(\frac{M(I)}{|I|}-1\right)\left(\frac{M(J)}{|J|}-1\right) \right)=\E\left(\frac{M(I)}{|I|} \frac{M(J)}{|J|}\right)-\E\left( \frac{M(I)}{|I|}\right)-\E\left( \frac{M(J)}{|J|}\right)+1.\nonumber
    \end{eqnarray}
By construction, $\omega$ is a gaussian process. Thus,
using Fubini's Theorem, we have:

\begin{eqnarray}
    \begin{cases}
        \E\left(\frac{M(I)}{|I|}\right)=\E\left(\frac{M(J)}{|J|}\right)=e^{\sum_{i=1}^d\frac{\left(a_i^2-a_i\right)\nu_i^2}{4}}\\
        \E\left(\frac{M(I)}{|I|} \frac{M(J)}{|J|}\right)=\frac{1}{\left|I\right|\left|J\right|}\int_{I}\int_{ J}e^{\sum_{i=1}^d\frac{\nu_i^2\left(a_i^2-a_i\right)}{2}+\sum_{i=1}^d a_i^2\text{Cov}\left(\omega^{i}_t,\omega^{i}_s\right)}{\rm d}t{\rm d}s 
    \end{cases}.
    \nonumber
\end{eqnarray}

As a result, we obtain:
\begin{eqnarray}
     \E\left(\left(\frac{M(I)}{|I|}-1\right)\left( \frac{M(J)}{|J|}-1\right) \right)=1+\sum_{i=1}^d\frac{\left(a_i^2-a_i\right)\nu_i^2}{2}+ 
     \sum_{i=1}^d\frac{a_i^2\lambda_i^2}{\left|I\right|\left|J\right|}\int_{I\times J}\text{Cov}\left(\frac{1}{\lambda_i}\omega^{i}_t,\frac{1}{\lambda_i}\omega^{i}_s)\right){\rm d}t{\rm d}s \nonumber \\
     -2-\sum_{i=1}^d\frac{\left(a_i^2-a_i\right)\nu_i^2}{2}
     +1+o\left(\left|\left|\LL\right|\right|^2\right). \nonumber\nonumber
    \end{eqnarray}
After some algebra, this leads to:
    \begin{eqnarray}
     \E\left(\left(\frac{M(I)}{|I|}-1\right)\left(\frac{M(J)}{|J|}-1\right) \right)=\sum_{i=1}^d\frac{a_i^2\lambda_i^2}{\left|I\right|\left|J\right|}\int_{I\times J}\text{Cov}\left(\frac{1}{\lambda_i}\omega^{i}_t,\frac{1}{\lambda_i}\omega^{i}_s\right){\rm d}t{\rm d}s+o\left(\left|\left|\LL\right|\right|^2\right),\nonumber
    \end{eqnarray}
Using Proposition C.1 in \cite{bacry2008log}, we have:
\begin{eqnarray}
         \E\left(\ln\left(\frac{M\left(I\right)}{|I|}\right)\ln\left(\frac{M\left(J\right)}{|J|}\right)\right)=\sum_{l=1}^d a_l^2 \lambda_{l}^2\E\left(\frac{\Upsilon_{l}(I)}{|I|}\frac{\Upsilon_{l}(J)}{|J|}\right)+o\left(\left|\left|\LL\right|\right|^2\right).\nonumber
\end{eqnarray}
In particular, we have:
\begin{eqnarray}
         \E\left(\ln\left(\frac{M\left(I\right)}{|I|}\right)\ln\left(\frac{M\left(I\right)}{|I|}\right)\right)=\sum_{l=1}^d a_l^2 \lambda_{l}^2\E\left(\frac{\Upsilon_{l}(I)}{|I|}\frac{\Upsilon_{l}(I)}{|I|}\right)+o\left(\left|\left|\LL\right|\right|^2\right),\nonumber
\end{eqnarray}
which leads to the desired result by substracting the two previous equalities.\\
    \item 
Given two arbitrary sub-intervals $I$ and $J$ of $K$, we have
\begin{eqnarray}
    \Sigma\left(I\right)\Sigma\left(J\right)=\sum_{1\leq l,s \leq d} b_l^2 b_s^2 M^l\left(I\right)M^s\left(J\right). \nonumber
\end{eqnarray}

For any $(l,s)\in \llbracket 1,d \rrbracket^2$, we have:

\begin{eqnarray}
\E \left( \frac{M^l(I)}{|I|}\frac{M^s(J)}{|J|} \right) = \int_{I \times J}\frac{dxdy}{|I||J|} \E \left( e^{\omega^l_x+\omega^s_y}\right)  . \nonumber
\end{eqnarray}

We have for any $(x,y)\in I\times I$:
\begin{eqnarray}
    \E\left(e^{\omega^l_x + \omega^s_y} \right) = e^{\mu_l + \mu_l + \frac{1}{2} \sum_{(s,l) \in \{l,s\}} \lambda_s \lambda_l \rho_{s,l}(|x-y|)}, \nonumber
\end{eqnarray}

where:
\begin{eqnarray}
    \rho_{i,j}\left(\tau\right)=
    \begin{cases*}
        \frac{1}{(H_i+H_j)(1-H_i-H_j)}\left(1-\left(\frac{\tau}{T}\right)^{H_i+H_j}\right) & \mbox{if}  i=j  \\
        0 & \mbox{otherwise} 
    \end{cases*}.\nonumber
\end{eqnarray}

Which gives after some algebra:
\begin{eqnarray}
    \E\left(e^{w_{H_{l},T}(x)+w_{H_{s},T}(y)}\right) = e^{\lambda_{l }\lambda_{s}\rho_{l, s}(|x-y|)}. \nonumber
\end{eqnarray}
By denoting the functional:
  $$\psi:\left(\lambda_{l},\lambda_{s}\right) \mapsto e^{\lambda_{l }\lambda_{s}\rho_{l ,s}(|x-y|)},  $$
we perform a first order Taylor expansion:
 \[\psi\left(\lambda_{l},\lambda_{s}\right)=\psi(0_{\R^2})+\left<\nabla_{\bm{\lambda}_d=0_{\R^2}}\psi,\bm{\lambda}\right>+\frac{1}{2}\left<Hess\left(\psi\right)|_{\bm{\lambda}=0_{\R^2}}\bm{\lambda},\bm{\lambda}\right>+ o\left(\left|\left|\bm{\lambda} \right|\right|^2\right),\]
Where 
$$ \bm{\lambda}= \begin{pmatrix}
\lambda_l \\
\lambda_s 
\end{pmatrix}
$$
and $Hess\left(.\right)$ is the Hessian operator.\\

We get:

 \[\psi\left(\lambda_{l},\lambda_{s}\right)=1+\lambda_{l}\lambda_{s}\rho_{l,s}(|x-y|)+ o\left(\left|\left|\bm{\lambda} \right|\right|^2\right).\]
Thus, we conclude that:

\begin{eqnarray}
    \E\left(\frac{M^l\left(I\right)}{|I|}\frac{M^s\left(J\right)}{|J|}\right)=
    \begin{cases*}
        1+ o\left(\left|\left|\bm{\lambda} \right|\right|^2\right)   \hspace{1.45cm} if \hspace{0.5mm} l \neq s \\
         1+\lambda_{l}^2\E\left(\frac{\Upsilon_{l}(I)}{|I|}\frac{\Upsilon_{l}(J)}{|J|}\right)+ o\left(\left|\left|\bm{\lambda} \right|\right|^2\right) \hspace{3mm} if \hspace{0.5mm}  l = s
    \end{cases*}.
    \nonumber
\end{eqnarray}

As a result, we obtain:
\begin{eqnarray}
          \E\left(\frac{\Sigma\left(I\right)}{|I|}\frac{\Sigma\left(J\right)}{|J|}\right)=\sum_{l=1}^d b_l^4 \left(1+\lambda_{l}^2\E\left(\frac{\Upsilon_{l}(I)}{|I|}\frac{\Upsilon_{l}(J)}{|J|}\right) \right)+\sum_{\substack{1 \leq l\neq s \leq d }}b_l^2b_s^2 +o\left(\left|\left|\LL \right|\right|^2\right) ,\nonumber
     \end{eqnarray}

which can be rewritten as:
\begin{eqnarray}
          \E\left(\frac{\Sigma\left(I\right)}{|I|}\frac{\Sigma\left(J\right)}{|J|}\right)=\sum_{l=1}^d b_l^4 \lambda_{l}^2\E\left(\frac{\Upsilon_{l}(I)}{|I|}\frac{\Upsilon_{l}(J)}{|J|}\right) +\left(\sum_{l=1}^d b_l^2\right)^2 +o\left(\left|\left|\LL\right|\right|^2\right).\nonumber
\end{eqnarray}
Thus, we obtain:
\begin{eqnarray}
         \E\left(\left(\frac{\Sigma\left(I\right)}{|I|}-1\right)\left(\frac{\Sigma\left(J\right)}{|J|}-1\right)\right)=\sum_{l=1}^d b_l^4 \lambda_{l}^2\E\left(\frac{\Upsilon_{l}(I)}{|I|}\frac{\Upsilon_{l}(J)}{|J|}\right)+\left(\sum_{l=1}^d b_l^2\right)^2 -2\sum_{l=1}^db_l^2 \nonumber \\
         +1 +o\left(\left|\left|\LL\right|\right|^2\right).\nonumber
\end{eqnarray}
Using the Proposition C.1 in \cite{bacry2008log}, we have:
\begin{eqnarray}
         \E\left(\ln\left(\frac{\Sigma\left(I\right)}{|I|}\right)\ln\left(\frac{\Sigma\left(J\right)}{|J|}\right)\right)=\sum_{l=1}^d b_l^4 \lambda_{l}^2\E\left(\frac{\Upsilon_{l}(I)}{|I|}\frac{\Upsilon_{l}(J)}{|J|}\right)+\left(\sum_{l=1}^d b_l^2\right)^2 -2\sum_{l=1}^db_l^2 \nonumber \\
         +1 +o\left(\left|\left|\LL\right|\right|^2\right).\nonumber
\end{eqnarray}
In particular, we have:
\begin{eqnarray}
         \E\left(\ln\left(\frac{\Sigma\left(I\right)}{|I|}\right)\ln\left(\frac{\Sigma\left(I\right)}{|I|}\right)\right)=\sum_{l=1}^d b_l^4 \lambda_{l}^2\E\left(\frac{\Upsilon_{l}(I)}{|I|}\frac{\Upsilon_{l}(I)}{|I|}\right)+\left(\sum_{l=1}^d b_l^2\right)^2 -2\sum_{l=1}^db_l^2 \nonumber \\
         +1 +o\left(\left|\left|\LL\right|\right|^2\right),\nonumber
\end{eqnarray}
which leads to the desired result by substracting the two previous equalities again.\\
\end{enumerate}
\tab\tab\tab\tab\tab\tab\tab\tab\tab\tab\tab\tab\tab\tab\tab\tab\tab\tab\tab\tab\tab\tab\tab\tab\tab\tab\tab\tab\tab\tab\tab\tab\tab\tab $\blacksquare$

Now, we proceed to the proof of Theorem \ref{theo:variancesumsmallintermit}.
\label{proof:variancesumsmallintermitequation}
We denote:
\begin{eqnarray}
\begin{cases}
    C^{log}\left(\Delta,\tau\right):=C^{log}([0,\Delta],[\tau,\tau+\Delta])\\
    C_{\Sigma}^{log}\left(\Delta,\tau\right):=C^{log}_{\Sigma}([0,\Delta],[\tau,\tau+\Delta])
\end{cases}
\end{eqnarray}

and for any $i\in \llbracket 1,d \rrbracket$:
\begin{eqnarray}
    C_{i}\left(\Delta,\tau\right):=C_i([0,\Delta],[\tau,\tau+\Delta]).
\end{eqnarray}
We denote as well for any $t\geq 0$: $M_{\Delta}(t):=M([t,t+\Delta])$. 
\begin{enumerate}[label=\alph*)]
 \item 
After some algebra, one has:
 \begin{eqnarray}
        W\left(\tau,\Delta,\LL\right)=2 \textnormal{var} \left(\ln\left(\frac{M_{\Delta}(\tau)}{\Delta}\right) \right)-2C^{log}\left(\Delta,\tau\right). \nonumber
    \end{eqnarray}
    Consequently, we get using Theorem \ref{thm:smallinterautocovinterval}:
\begin{eqnarray}
        W\left(\tau,\Delta,\LL\right)=\sum_{i=1}^d\frac{2a_i^2 \lambda_{i}^2}{\Delta^2}\left(C_{i}\left(\Delta,0\right)-C_{i}\left(\Delta,\tau\right)\right) +o\left(\left|\left|\LL\right|\right|^2\right).\nonumber
\end{eqnarray}
After some algebra, we obtain:
\begin{eqnarray}
        W\left(\tau,\Delta,\LL\right)=\sum_{l=1}^d a_{l}^2\lambda_{l}^2\frac{|\tau+\Delta|^{2H_l+2}-2\tau^{2H_l+2}+|\tau-\Delta|^{2H_l+2}-2|\Delta|^{2H_l+2}}{\Delta^2H_lT^{2H_l}(1-(2H_l)^2)(2H_l+2)}+\nonumber o\left(\left|\left|\LL\right|\right|^2\right).\nonumber
\end{eqnarray}

 Consequently, we get:
 \begin{eqnarray}
        W\left(\tau,\Delta,\LL\right)=\sum_{l=1}^d a_{l}^2\lambda_{l}^2\left(\frac{\tau}{T}\right)^{2H_l}g_{H_l}\left(\frac{\Delta}{\tau}\right) +o\left(\left|\left|\LL\right|\right|^2\right),\nonumber
\end{eqnarray}
where the function $g_{H}(.)$ is defined in \eqref{eq:defg}. \\\\ 
This leads to:
\begin{eqnarray}
    \mathcal{W}\left(\tau,\Delta,\LL\right)=\sum_{l=1}^d a_{l}^2 \lambda_{l}^2\left(\frac{\tau}{T}\right)^{2H_l}g_{H_l}\left(\frac{\Delta}{\tau}\right). \nonumber
\end{eqnarray}

    \item 
 This part is analogous to the previous one. We have:
 \begin{eqnarray}
        V\left(\tau,\Delta,\LL\right)=2\textnormal{var}\left(\ln\left(\frac{\Sigma_{\Delta}(\tau)}{\Delta}\right) \right)-2C_{X}^{log}\left(\Delta,\tau\right). \nonumber
    \end{eqnarray}
    Consequently, we get using Theorem \ref{thm:smallinterautocovinterval}:
\begin{eqnarray}
        V\left(\tau,\Delta,\LL\right)=\sum_{i=1}^d\frac{2b_i^4 \lambda_{i}^2}{\Delta^2}\left(C_{i}\left(\Delta,0\right)-C_{i}\left(\Delta,\tau\right)\right) +o\left(\left|\left|\LL\right|\right|^2\right).\nonumber
\end{eqnarray}
After some algebra, we obtain:
 \begin{eqnarray}
        V\left(\tau,\Delta,\LL\right)=\sum_{l=1}^d b_l^4 \lambda_{l}^2\left(\frac{\tau}{T}\right)^{2H_l}g_{H_l}\left(\frac{\Delta}{\tau}\right) +o\left(\left|\left|\LL\right|\right|^2\right),\nonumber
\end{eqnarray}
where the function $g_{H}(z)$ is defined in \eqref{eq:defg}. \\\\ 
This leads to:
\begin{eqnarray}
    \mathcal{V}\left(\tau,\Delta,\LL\right)=\sum_{l=1}^d b_l^4 \lambda_{l}^2\left(\frac{\tau}{T}\right)^{2H_l}g_{H_l}\left(\frac{\Delta}{\tau}\right). \nonumber
\end{eqnarray}
\end{enumerate}

\tab\tab\tab\tab\tab\tab\tab\tab\tab\tab\tab\tab\tab\tab\tab\tab\tab\tab\tab\tab\tab\tab\tab\tab\tab\tab\tab\tab\tab\tab\tab\tab\tab\tab $\blacksquare$

\renewcommand{\thesubsection}{A.4}
\section{Hurst exponent estimation from moment scaling}
\label{subsubsec:hurstindentification}
In order to measure the Hurst exponent $H_X$ associated with some given process $X(t)$ we can refer to the method
used in \cite{gatheral2018volatility} that simply consists in estimating the scaling exponent of various moment
of the increments of the process $\delta_\tau X(t) = X(t+\tau)-X(t)$. 
In the case when $X(t) = \ln \Sigma_{\Delta}(t ) $, the increment process is simply $\delta_{\tau}\ln \Sigma_{\Delta}(t )  = \ln \Sigma_{\Delta}(t + \tau ) - \ln \Sigma_{\Delta}(t )$. We have seen that, at the first order in $\left|\left|\LL\right|\right|^2$, its moments $m(q, \tau, \Delta)= \E \left( \left | \delta_{\tau} \ln \Sigma_{\Delta}(t) \right|^q \right)$
correspond to the moments a Gaussian process and therefore one has:
\begin{eqnarray}
    m(q, \tau, \Delta) \underset{\left|\left|\LL\right|\right|^2 \ll 1}= \pi^{-\frac{1}{2}} 2^{\frac{q}{2}}\Gamma\left(q +\frac{1}{2}\right) \mathcal{V}\left(\tau,\Delta,\LL\right)^{\frac{q}{2}},
\end{eqnarray}
where $ \mathcal{V}\left(\tau,\Delta,\LL\right)$ is defined in Section \ref{sec:SIA} and
$\Gamma \left( x \right) = \int\limits_0^{+\infty} {s^{x - 1} e^{ - s} {\rm d}s},x>0$ while
$\underset{\left|\left|\LL\right|\right|^2 \ll 1}=$ indicates that equality holds in the first order when $\left|\left|\LL\right|\right|^2 \rightarrow 0$). \\\\
There, if As there exists a constant $C>0$ such that:
\begin{eqnarray}
     \mathcal{V}\left(\tau,\Delta,\LL\right) = C\tau^{2H_X}, \nonumber
\end{eqnarray}
$H_X$ can be identified from the slope of the linear regression of $\ln\left(m(q, \tau, \Delta) \right)$ against $\ln(\tau)$.\\\\
Besides, if for any $l\in \llbracket 1,d \rrbracket,H_l\neq 0 $, we have the following Taylor expansion:
 \begin{eqnarray}
 \label{eq:gapprox}
        g_{H_l}\left(z\right)=\frac{1}{H_{l}(1-2H_{l})}+ o\left(z^{2H_{l}}\right)    
\end{eqnarray}

Consequently, if one supposes for instance that:
\[
    \begin{cases}
     \exists c > 0, \forall i \in \llbracket 1, d \rrbracket \text{ such that } b_i^2 \lambda_i = c, \\
     H_1 \ll H_2 \ll \cdots \ll H_d
    \end{cases}
\]

The Eq.~\eqref{eq:variancesumsmallintermitequation} can be simplified as follows in the high frequency limit ($\Delta \ll \tau$):
\begin{eqnarray}
        \mathcal{V}\left(\tau,\Delta,\LL\right)\approx \frac{\tau^{2H_1}}{H_{1}(1-2H_{1})}.\nonumber
\end{eqnarray}
Leading to: $$H_X \approx H_1.$$

 The latter result infers that in the high frequency limit, we observe only the regularity of the process $X^{i}$ with the smallest Hurst exponent through the aggregated process $X$.\\\\

\renewcommand{\thesubsection}{A.5}
\section{Proof of Eq.~\eqref{eq:ineqr_0H}}
\label{app:proofineqr_0H}
By definition, one has for any $z \in ]0,1[$:
\begin{eqnarray}
r_{0,H}\left(z\right)=\frac{g_0(z)}{g_{H}(z)}=H(1-(2H)^2)(2H+2)\frac{ \frac{(z + 1)^2 \ln(z + 1) + (-1 + z)^2 \ln(1 - z) - 2z^2 \ln(z)}{z^2}}{\frac{|1+z|^{2H+2}-2|z|^{2H+2}+|1-z|^{2H+2}-2}{z^2}}.\nonumber
\end{eqnarray}
For any $H \in \left]0,\frac{1}{2}\right[$ , $h_{H}:z\mapsto \frac{|1+z|^{2H+2}-2|z|^{2H+2}+|1-z|^{2H+2}-2}{z^2}$ is a decreasing function in $]0,1[$. As a result:
\begin{eqnarray}
\forall z\in ]0,1[,\tab h_{H}(z) >h_{H}(1) = 4(2^{2H}-1)\nonumber
\end{eqnarray}
On the other hand, one has:
\begin{eqnarray}
    \forall z\in ]0,1[,\tab
    \begin{cases}
        \ln\left(1+z\right) \leq z-\frac{z^2}{2}+\frac{z^3}{3}\\
        \ln\left(1-z\right) \leq -z-\frac{z^2}{2}-\frac{z^3}{3}
    \end{cases}
    \nonumber
\end{eqnarray}
For any $z\in ]0,1[$, one has after some algebra:
\begin{eqnarray}
    g_{0}(z)\leq 3-\frac{z^2}{2}-2\ln(z) \leq 3-2\ln(z) \nonumber
\end{eqnarray}
Consequently, the following holds:
\begin{eqnarray}
    r_{0,H}(z)\leq \frac{H(1-2H)(1+2H)(H+1)}{2(2^{2H}-1)}\left(3-2\ln(z)\right)  \nonumber
\end{eqnarray}
Which ends the proof.\\\tab\tab\tab\tab\tab\tab\tab\tab\tab\tab\tab\tab\tab\tab\tab\tab\tab\tab\tab\tab\tab\tab\tab\tab\tab\tab\tab\tab\tab\tab\tab\tab\tab\tab $\blacksquare$

\renewcommand{\thesubsection}{A.6}
\section{Proof of Proposition \ref{prop:qvfactorintermsofqvassets}}
\label{app:proofqvfactorintermsofqvassets}
Let us define the increment operator $\delta_\tau$ acting on an arbitrary stochastic process $x_t$ as follows:
\begin{eqnarray}
    \forall (\tau,t)\in K^2, \tab \delta_{\tau} x_t:=x_{t+\tau}-x_t. \nonumber
\end{eqnarray}
Starting from the N-SfFM in  Eq.~\eqref{eq:NFM_definition}, we have for any $\tau\in K$:
\begin{equation}
\forall i\in \llbracket 1,N_s \rrbracket,\quad
   \delta_{\tau} x^{i}_t=\beta_i\delta_{\tau} f_t+\delta_{\tau} e^{i}_t,
    \nonumber
\end{equation}
where:
\begin{eqnarray}
    \begin{cases*}
        \delta_{\tau} x^{i}_t=\int_t^{t+\tau}{\rm d}x^{i}_s\\
        \delta_{\tau} f_t =\int_t^{t+\tau}e^{\frac{\Omega_{s}}{2}}{\rm d}W_s\\
        \delta_{\tau} e^{i}_t =\int_t^{t+\tau}e^{\frac{\gamma_i\Omega_{s}+\omega^{i}_{s}}{2}}{\rm d}B^{i}_s
    \end{cases*}.
    \nonumber
\end{eqnarray}
We consider the two optimal least square solutions:
\begin{eqnarray}
\label{eq:olsfactorincrement}
\forall t\in K,\quad
\begin{cases}
    \widehat{\delta_{\tau}f}_{t}:=\underset{f\in \R}\argmin  \left|\left|\delta_{\tau} X_t-\beta  f\right|\right|\\
    \widehat{f}_{t}:=\underset{f\in \R}\argmin  \left|\left| X_t-\beta f\right|\right|
\end{cases},
\end{eqnarray}
where $\left|\left| .\right|\right|$ is the $\mathcal{L}^2$ norm in $\R^{N_s}$.\\\\
It is straightforward to see that the ordinary least squares solutions of Eq.~\eqref{eq:olsfactor} are linked as follows.
\begin{eqnarray}
\label{eq:lemhatfactor}
        \widehat{\delta_{\tau}f}_{t}=\delta_{\tau}\widehat{f}_{t}.
    \end{eqnarray}
We have timewise the following ordinary least square solution:
    \begin{eqnarray}
        \forall t\in K, \tab \quad\widehat{\delta_{\tau}f}_t = \frac{\sum_{i=1}^N\beta_i\delta_{\tau}x^{i}_t}{\sum_{i=1}^N\beta_i^2}. \nonumber
    \end{eqnarray}
    For the sake of simplicity, we denote: $h:=\frac{1}{\sum_{i=1}^N\beta_i^2}$.\\\\
As a result, we obtain:
\begin{eqnarray}
    \widehat{\delta_{\tau}f}_{t}^2 = h^2\sum_{k^{'},k^{''}=1}^N\beta_{k^{'}}\beta_{k^{''}} \delta_{\tau}x^{k^{'}}_t\delta_{\tau}x^{k^{''}}_t.\nonumber
\end{eqnarray}
Given that for any $ i\in \llbracket 1,N \rrbracket,\quad\delta_{\tau}x^{i}_t=\beta_{i}\delta_{\tau}f_t+\delta_{\tau}e^{i}_t$, we have:
\begin{eqnarray}
    \widehat{\delta_{\tau}f}_{t}^2 = \sum_{k^{'}=1}^N h^2\beta_{k^{'}}^2\left(\delta_{\tau}x^{k^{'}}_{t}\right)^2+\sum_{k^{'}\neq k^{''}=1}^N h^2\beta_{k^{'}}\beta_{k^{''}} \left(\beta_{k^{'}}\delta_{\tau}f_t+\delta_{\tau}e^{k^{'}}_t\right)\left(\beta_{k^{''}}\delta_{\tau}f_t+\delta_{\tau}e^{k^{''}}_t\right).\nonumber
\end{eqnarray}
We obtain:
\begin{eqnarray}
    \widehat{\delta_{\tau}f}_{t}^2 = \sum_{k^{'}=1}^Nh^2\beta_{k^{'}}^2\left(\delta_{\tau}x^{k^{'}}_{t}\right)^2+\left(\sum_{k^{'}\neq k^{''}=1}^Nh^2\left(\beta_{k^{'}}\beta_{k^{''}} \right)^2\right)\left(\delta_{\tau}f_{t}\right)^2 \nonumber \\
    \sum_{k^{'}\neq k^{''}=1}^N h^2\beta_{k^{'}}^2\beta_{k^{''}}\delta_{\tau}f_t\delta_{\tau}e^{k^{''}}_t +
    \sum_{k^{'}\neq k^{''}=1}^N h^2\beta_{k^{'}}\beta_{k^{''}}^2\delta_{\tau}f_t\delta_{\tau}e^{k^{'}}_t+\sum_{k^{'}\neq k^{''}=1}^N h^2\beta_{k^{'}}\beta_{k^{''}}\delta_{\tau}e^{k^{'}}_t\delta_{\tau}e^{k^{''}}_t .\nonumber
\end{eqnarray}

As the previous equality is valid for any $t$ and $\tau$, considering a subdivision $\left(t_k\right)_{k\in \llbracket 1,n \rrbracket}$ with time step $\theta:=\frac{t_n-t_0}{n}$ of an arbitrary sub interval $I\subset K$.\\\\
By definition, we have:
\begin{eqnarray}
\forall (i,j)\in \llbracket 1,N \rrbracket^2,
    \begin{cases}
         \sum_{l=1}^n \left(\delta_{t_l}x^{i}_{t_{l-1}}\right)^2 \xrightarrow[\theta\rightarrow 0]{\Prob} \left<x^{i}\right>_I\\
         \sum_{l=1}^n \left(\delta_{t_l}f_{t_{l-1}}\right)^2 \xrightarrow[\theta\rightarrow 0]{\Prob} \left<f\right>_I\\
         \sum_{l=1}^n \delta_{t_l}f_{t_{l-1}}\delta_{t_l}e^{i}_{t_{l-1}}\xrightarrow[\theta\rightarrow 0]{\Prob} \left<f,e^{i}\right>_I \\
         \sum_{l=1}^n \delta_{t_l}e^{i}_{t_{l-1}}\delta_{t_l}e^j_{t_{l-1}}\xrightarrow[\theta\rightarrow 0]{\Prob} \left<e^{i},e^j\right>_I 
    \end{cases}.
    \nonumber
\end{eqnarray}

Given that $B$ and $B^j$ are independent from $B^{i}$ for any $i\neq j\in \llbracket 1,N \rrbracket^2$, we have that:
\begin{eqnarray}
    \forall i\neq j\in \llbracket 1,N \rrbracket,\quad\tab  \left<f,e^{i}\right>_I=0,\tab \left<e^{i},e^j\right>_I=0 .\nonumber
\end{eqnarray}
This means that the right hand side of the previous equation converges in probability as $\theta \rightarrow 0$ to: $\sum_{k^{'}=1}^Nh^2\beta_{k^{'}}^2\left<x^{k^{'}}\right>_I+\left(\sum_{k^{'}\neq k^{''}=1}^Nh^2\left(\beta_{k^{'}}\beta_{k^{''}} \right)^2\right)\left<f\right>_I $.\\\\
On the other hand, we have using Eq.~\eqref{eq:lemhatfactor}: 
\begin{eqnarray}
    \sum_{l=1}^n \left(\widehat{\delta_{t_l} f}_{t_{l-1}}\right)^2 \xrightarrow[\theta\rightarrow 0]{\Prob} \left<\hat{f}\right>_I ,\nonumber
\end{eqnarray}
and by the almost sure uniqueness of the limit in probability, we conclude that:

\begin{eqnarray}
\label{refequationqv}
    \left<\widehat{f}\right>_I= \sum_{k^{'}=1}^Nh^2\beta_{k^{'}}^2\left<x^{k^{'}}\right>_I+\left(\sum_{k^{'}\neq k^{''}=1}^Nh^2\left(\beta_{k^{'}}\beta_{k^{''}} \right)^2\right)\left<f\right>_I .
\end{eqnarray}
On the other hand, starting from this identity:
\begin{eqnarray}
    \forall (t,\tau)\in \R_{+}^2 ,   \quad\widehat{\delta_{\tau}f}_t = \delta_{\tau}f_t+\sum_{i=1}^Np_i\delta_{\tau}e^{i}_t.\nonumber
    \end{eqnarray}
where $p_i=\frac{\beta_i\sigma_i}{\sum_{i=1}^N\beta_i^2}$ for $i\in\llbracket1,N\rrbracket$.\\
Using the same convergence in probability arguments, we end up with:
\begin{eqnarray}
       \left<\hat{f}\right>_I = \left<f\right>_I+\sum_{i=1}^Np_i^2\left<e^{i}\right>_I, \tab a.s.\nonumber
    \end{eqnarray}
Using the strong law of large numbers, we have that:
\begin{eqnarray}
       \left<\hat{f}\right>_I = \left<f\right>_I+\underset{N\rightarrow \infty} o\left(1\right), a.s.\nonumber
    \end{eqnarray}

By injecting the latter in Eq.~\eqref{refequationqv}, we obtain after some algebra that:
\begin{eqnarray}
    \left<\hat{f}\right>_I \underset{N\rightarrow\infty}\sim  \frac{1}{h^{-2}-\left(\sum_{k^{'}\neq k^{''}=1}^N\left(\beta_{k^{'}}\beta_{k^{''}} \right)^2\right)}\sum_{k^{'}=1}^N\beta_{k^{'}}^2\left<x^{k^{'}}\right>_I, \tab a.s, \nonumber
\end{eqnarray}
from which the claimed result derives.
\\
\tab\tab\tab\tab\tab\tab\tab\tab\tab\tab\tab\tab\tab\tab\tab\tab\tab\tab\tab\tab\tab\tab\tab\tab\tab\tab\tab\tab\tab\tab\tab\tab\tab\tab $\blacksquare$
\label{app:proof}
We dedicate the end of this appendix in discussing Eq.~\eqref{eq:qvfactor}. Quantitatively speaking, the assumption over the exposure vector $\bf{\beta}$ in Proposition \ref{prop:qvfactorintermsofqvassets} can be restricted to considering its entries with absolute continuous entries with respect to the Lebesgue measure as no entry of $\beta$ in that case will charge an atom, in particular 0. Thus, the denominator in Eq.~\eqref{eq:qvfactor} will almost surely not vanish.\\
\renewcommand{\thesubsection}{H.1}
\section{Numerical Miscellanea}
\label{app:miscnumerics}
\subsection{Gaussian trick}
\label{app:Gaussian_trick}

We consider a noisy Gaussian random variable: 
\begin{eqnarray}
    Y=X+\epsilon, \nonumber
\end{eqnarray}
where $X\sim \mathcal{N}(0,1)$ and $\epsilon$ a random noise.
\\\\

To better reconstruct the distribution of $X$ from $Y$, we consider $\Tilde{X}$ as a proxy of $X$ defined as follows:
\begin{eqnarray}
    \Tilde{X}:=F^{-1}\left(F_y\left(Y\right)\right),
\end{eqnarray}
where:
\begin{itemize}
    \item $F(x)=\frac{1}{\sqrt{2\pi}}\int_{-\infty}^x e^{-\frac{t^2}{2}}{\rm d}t$ the distribution function of a standard Gaussian random variable;
    \item $F_y(x)=\Prob\left(Y\leq x\right)$ the distribution function of $Y$.
\end{itemize}
Here is an example of such a reconstruction using the empirical distribution function of $Y$ as a proxy of $F_y$.
\begin{figure}[H]
    \centering
    \includegraphics[width=0.6\linewidth]{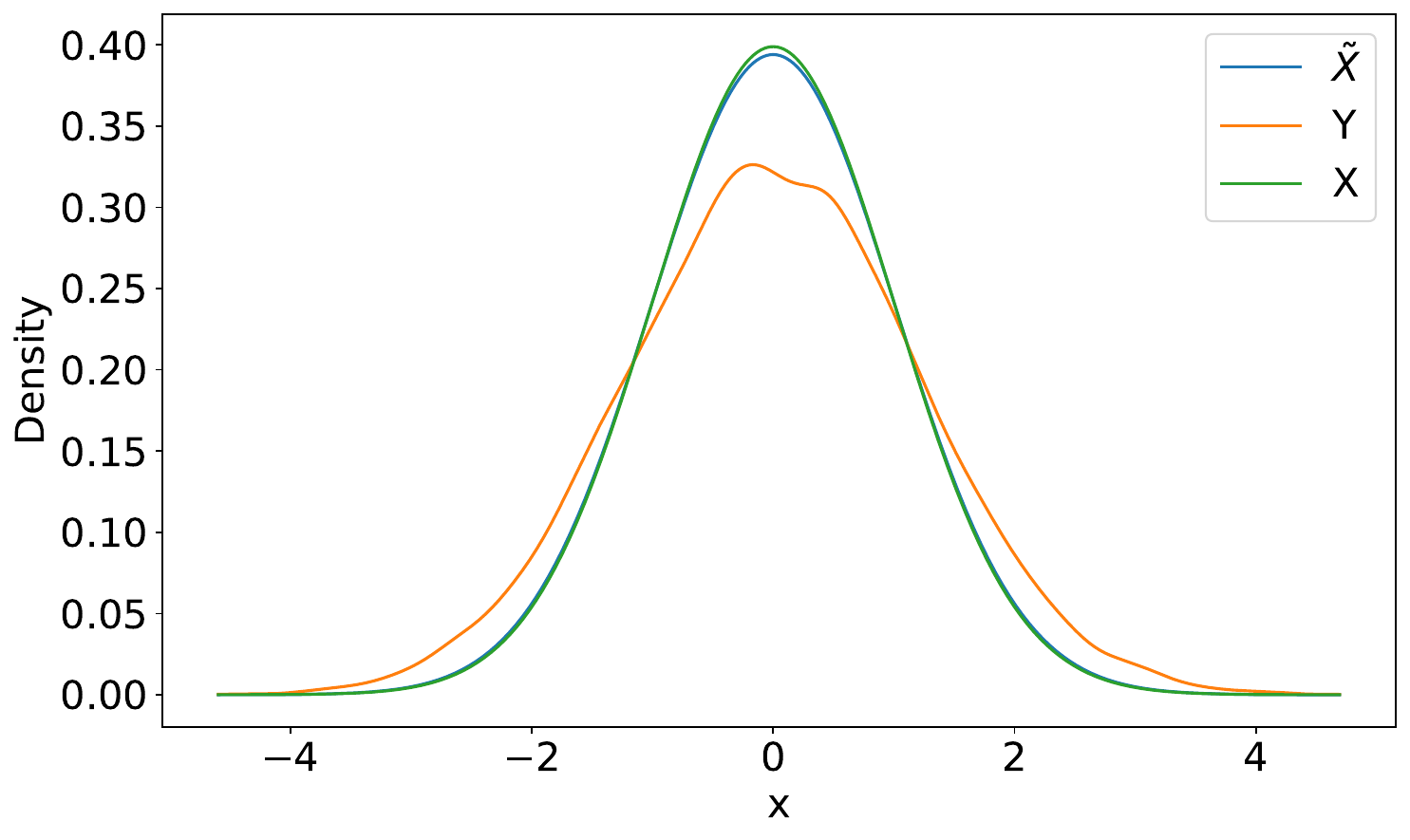} 
        \label{fig:subfig1}
    \caption{The true Gaussian density $X$ (\textbf{green}), the perturbated Gaussian $Y$ (\textbf{orange}) and the Gaussianized random variable's density $\Tilde{X}$ (\textbf{blue}), for both of which, we perform a Gaussian kernel density estimator. Here we considered a sample of $10^4$ drawn from $\mathcal{N}(0,1)$ as $X$ and a noise $\epsilon$ drawn from $\mathcal{N}(0,0.7)$.}
    \label{fig:main}
\end{figure}
\renewcommand{\thesubsection}{G.2}

\renewcommand{\thesubsection}{H.2}
\subsection{Empirical distributions of $\beta$ and $\sigma$ entries}
\label{app:betasigmaemiricaldistrib}
Using empirical daily data from YahooFinance as in Section \ref{sec:EstimationRoughnessIdiosyncratic}  spanning from 01/01/1994 to 01/01/2026 for 243 US stocks of the S\&P500 throughout the entire period, we find that a leptokurtic distribution fits well the observed histogram of the empirical distribution of the exposures. The latter exposures are calibrated following Step 1 in Section \ref{sec:calibrationmethod} (see Eq.~\eqref{eq:betaestimation}).
\begin{figure}[H]
    \centering
    \includegraphics[width=0.75\linewidth]{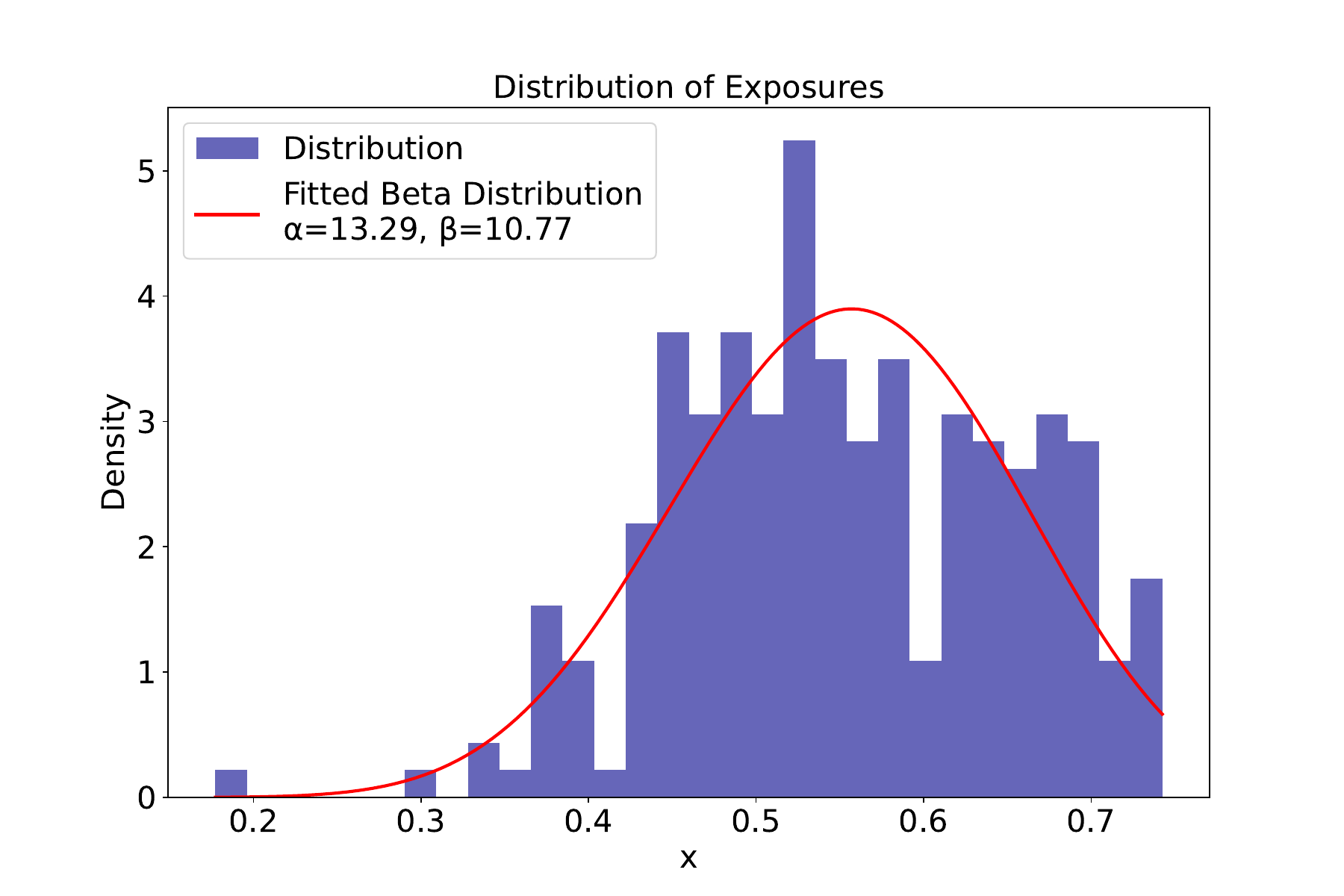} 
        \label{fig:subfig1}
    \caption{The empirical distribution of the $\beta$ entries (\textbf{blue}) together with the fitted beta distribution (\textbf{red}).}
    \label{fig:distributionbetas}
\end{figure}
On the other hand, to build intuition for the empirical distributions of $\sigma$ entries, we conduct a similar analysis by regressing the asset returns of the S\&P500 stocks spanning the same period and retrieve the calibrated entries of $\sigma$ via Eq.~\eqref{eq:qvresiduals} through their identification as the square root of the variance of the residuals.
It turns out that they also fit a leptokurtic distribution.
\begin{figure}[H]
    \centering
    \includegraphics[width=0.75\linewidth]{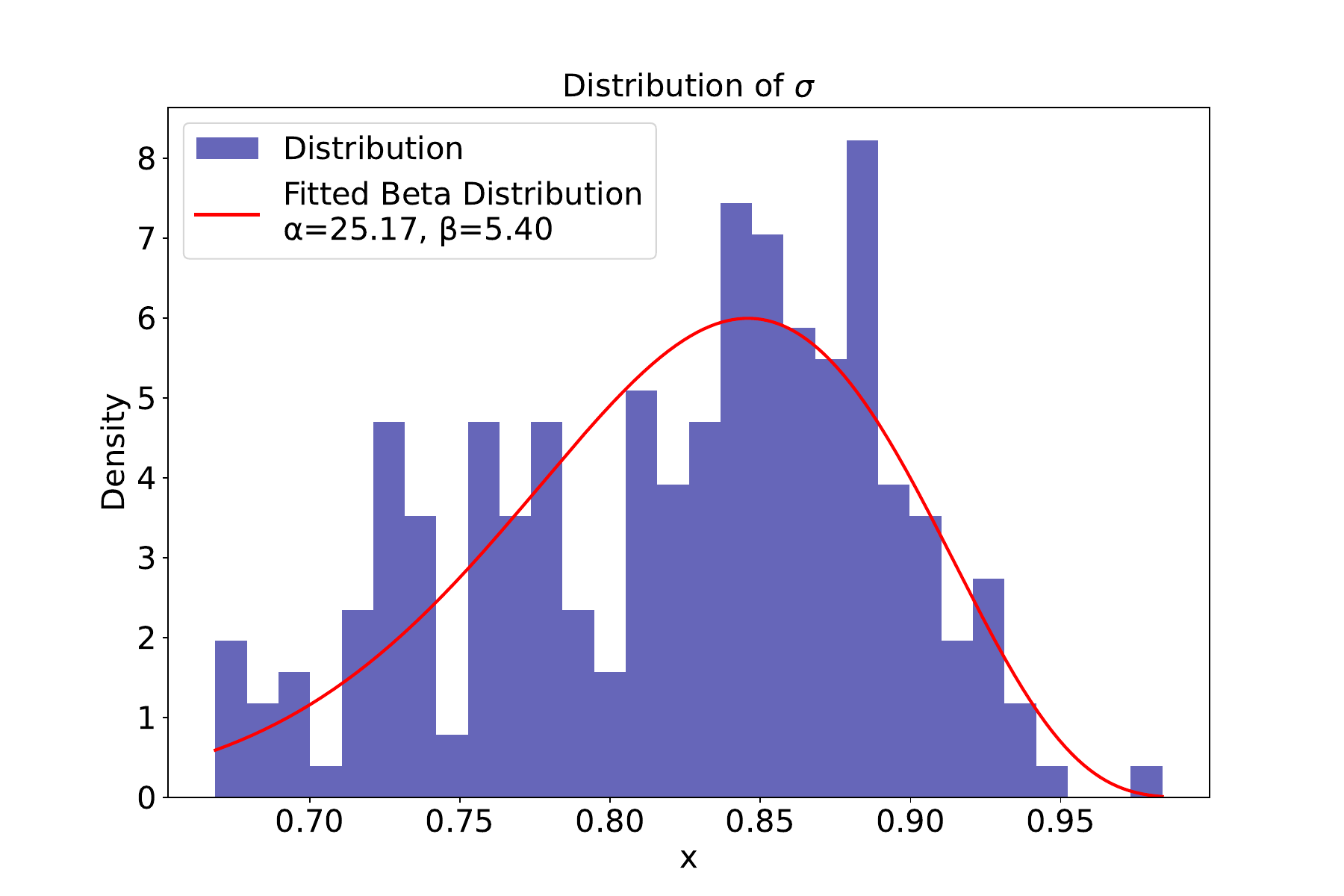} 
        \label{fig:subfig1}
    \caption{The empirical distribution of the $\sigma$ entries (\textbf{blue}) together with the fitted beta distribution (\textbf{red}).}
    \label{fig:distributionsigmas}
\end{figure}

The following figure displays the empirical distribution of the ratio's entries of $\frac{\beta}{\sigma}$ across all the considered assets. This aims to better illustrate empirically the condition of Eq.~\eqref{eq:beta_criterion}.

\begin{figure}[H]
    \centering
    \includegraphics[width=0.75\linewidth]{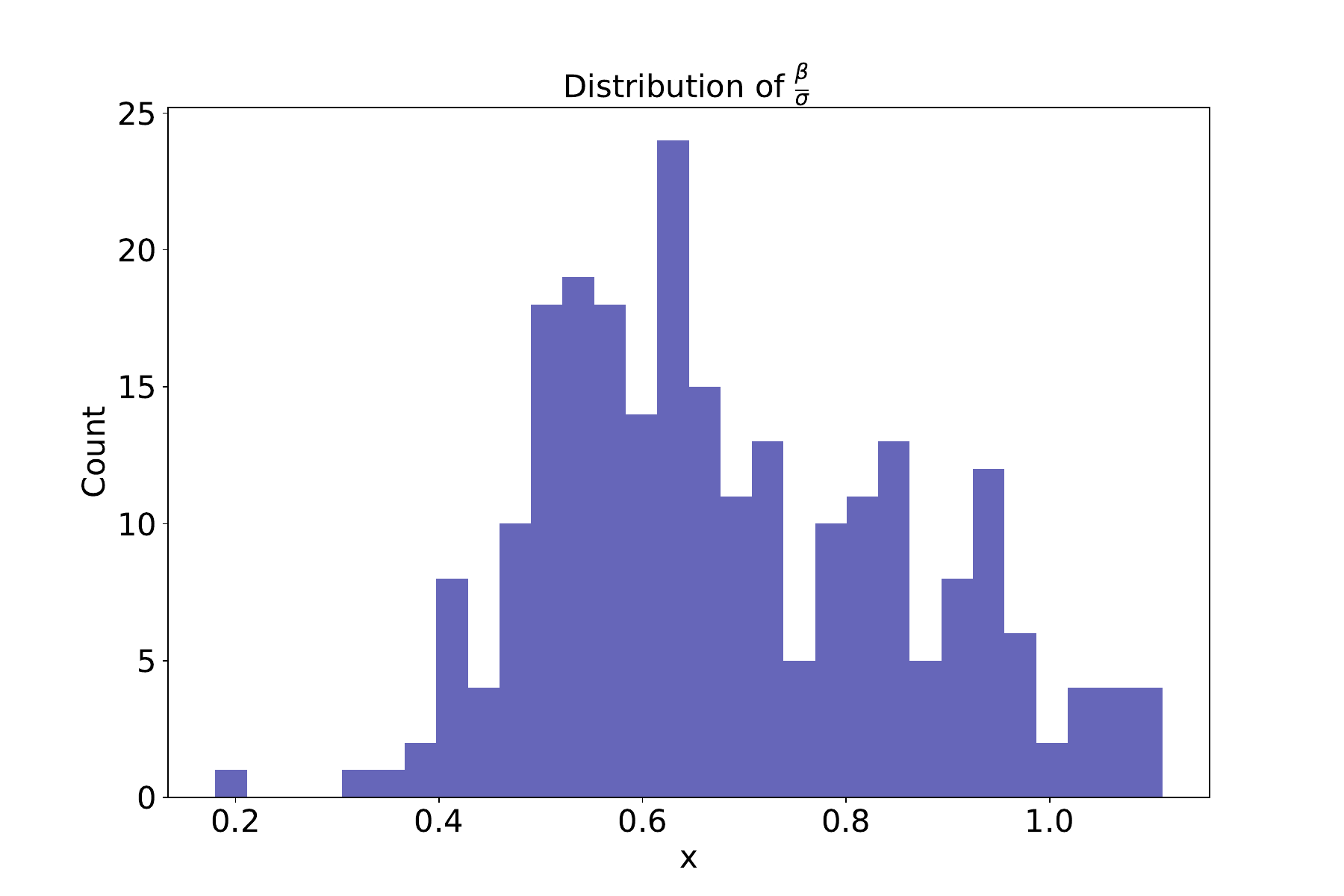} 
        \label{fig:subfig1}
    \caption{The empirical distribution of the $\frac{\beta}{\sigma}$ entries (\textbf{blue}).}
    \label{fig:distributionbetasoversigmas}
\end{figure}

The latter figure shows values of the ratio being mainly concentrated around $(0.4-0.9)$, meaning that the condition of Eq.~\eqref{eq:beta_criterion} is widely satisfied for the considered single stocks of the S\&P500. The few exceptions we observe where the ratio $\frac{\beta_i}{\sigma_i}$ exceeds $1$ correspond likely to stocks having a behavior similar to an index as emphasized in Section \ref{subSection:Hurstexponentsofsinglestocksandstockindexes}.

\section{Yahoo Finance S\&P500 asset nomenclature of Figure \ref{fig:omegaifrommktdata}}
\label{app:assetnomenclature}

\begin{longtable}{|c|l|l|}
\hline
\textbf{Ticker} & \textbf{Company Name} & \textbf{Sector} \\
\hline
\endfirsthead

\hline
\textbf{Ticker} & \textbf{Company Name} & \textbf{Sector} \\
\hline
\endhead

\hline
\endfoot

\hline
\endlastfoot

MO   & Altria Group & Consumer Staples \\
AMAT & Applied Materials & Information Technology \\
HBAN & Huntington Bancshares & Financials \\
VLO  & Valero Energy & Energy \\
SPG  & Simon Property Group & Real Estate \\
ADI  & Analog Devices & Information Technology \\
AMGN & Amgen & Health Care \\
LRCX & Lam Research & Information Technology \\
AEP  & American Electric Power & Utilities \\

\end{longtable}

\end{appendices}

\end{document}